\RequirePackage{lineno}
\documentclass[aps,prl,superscriptaddress,preprint]{revtex4-1}

\usepackage{graphicx,amsmath}
\usepackage{color}

\usepackage{booktabs}
\usepackage[flushleft]{threeparttable}

\usepackage{amssymb}
\usepackage{amsmath}
\usepackage{float}
\usepackage{psfrag}
\usepackage{siunitx}
\usepackage{ulem}

\def\be{\begin{equation}}
\def\ee{\end{equation}}
\def\ba{\begin{eqnarray}}
\def\ea{\end{eqnarray}}

\renewcommand{\Re}{\operatorname{Re}}
\newcommand{\Ra}{\operatorname{Ra}}
\newcommand{\Ta}{\operatorname{Ta}}

\newcommand{\Nu}{\operatorname{Nu}_{\omega}}

\newcommand{\Ro}{\operatorname{Ro}}

\begin{document}

\title{Wall roughness induces asymptotic ultimate turbulence}
\author{Xiaojue Zhu } \thanks{These authors contributed equally to this work}
\affiliation{Physics of Fluids Group and Max Planck Center Twente for Complex Fluid Dynamics, MESA+ Institute and J. M. Burgers Centre for Fluid Dynamics, University of Twente, P.O. Box 217, 7500AE Enschede, The Netherlands}

\author{Ruben A. Verschoof} \thanks{These authors contributed equally to this work}
\affiliation{Physics of Fluids Group and Max Planck Center Twente  for Complex Fluid Dynamics, MESA+ Institute and J. M. Burgers Centre for Fluid Dynamics, University of Twente, P.O. Box 217, 7500AE Enschede, The Netherlands}

\author{Dennis Bakhuis}
\affiliation{Physics of Fluids Group and Max Planck Center Twente for Complex Fluid Dynamics, MESA+ Institute and J. M. Burgers Centre for Fluid Dynamics, University of Twente, P.O. Box 217, 7500AE Enschede, The Netherlands}

\author{Sander G. Huisman}
\affiliation{Physics of Fluids Group and Max Planck Center Twente for Complex Fluid Dynamics, MESA+ Institute and J. M. Burgers Centre for Fluid Dynamics, University of Twente, P.O. Box 217, 7500AE Enschede, The Netherlands}

\author{Roberto Verzicco }
\affiliation{Dipartimento di Ingegneria Industriale, University of Rome ``Tor Vergata", Via del Politecnico 1, Roma 00133, Italy}
\affiliation{Physics of Fluids Group and Max Planck Center Twente for Complex Fluid Dynamics, MESA+ Institute and J. M. Burgers Centre for Fluid Dynamics, University of Twente, P.O. Box 217, 7500AE Enschede, The Netherlands}

\author{Chao Sun}
\email{chaosun@tsinghua.edu.cn}
\affiliation{Center for Combustion Energy and Department of Thermal Engineering, Tsinghua University, 100084 Beijing, China}
\affiliation{Physics of Fluids Group and Max Planck Center Twente for Complex Fluid Dynamics, MESA+ Institute and J. M. Burgers Centre for Fluid Dynamics, University of Twente, P.O. Box 217, 7500AE Enschede, The Netherlands}

\author{Detlef Lohse}
\email{d.lohse@utwente.nl}
\affiliation{Physics of Fluids Group and Max Planck Center Twente for Complex Fluid Dynamics, MESA+ Institute and J. M. Burgers Centre for Fluid Dynamics, University of Twente, P.O. Box 217, 7500AE Enschede, The Netherlands}
\affiliation{Center for Combustion Energy and Department of Thermal Engineering, Tsinghua University, 100084 Beijing, China}
\affiliation{Max Planck Institute for Dynamics and Self-Organization, 37077 G\"ottingen, Germany}


\begin{abstract} 
Turbulence is omnipresent in Nature and technology, governing the transport of heat, mass, and momentum on multiple scales. For real-world applications of wall-bounded turbulence, the underlying surfaces are virtually always rough; yet characterizing and understanding the effects of wall roughness for turbulence remains a challenge, especially for rotating and thermally driven turbulence. By combining extensive experiments and numerical simulations, here, taking as example the paradigmatic Taylor-Couette system (the closed flow between two independently rotating coaxial cylinders), we show how wall roughness greatly enhances the overall transport properties and the corresponding scaling exponents. If only one of the walls is rough, we reveal that the bulk velocity is slaved to the rough side, due to the much stronger coupling to that wall by the detaching flow structures. If both walls are rough, the viscosity dependence is thoroughly eliminated in the boundary layers and we thus achieve asymptotic ultimate turbulence, i.e. the upper limit of transport, whose existence had been predicted by Robert \ Kraichnan in 1962 (Phys. Fluids {\bf 5}, 1374 (1962)) and in which the scalings laws can be extrapolated to arbitrarily large Reynolds numbers.

\end{abstract}


\maketitle



\section{Introduction} 

While the vast majority of studies on wall-bounded turbulence assumes smooth walls, in engineering applications and even more so in nature, flow boundaries are in general rough, leading to a coupling of the small roughness scale to the much larger outer length scale of the turbulent flow. This holds for the atmospheric boundary layer over canopy or buildings, for geophysical flows, in oceanography, but also for many industrial flows such as pipe flow, for which the presumably most famous (though controversial) study on roughness was performed \cite{nik33}. For more recent works on the effect of wall roughness in (pipe or channel) turbulence we refer to various studies \cite{hul13,cha15,chu15,squ16}, reviews \cite{jim04,fla14}, or textbooks \cite{pop00,sch00}.


Rather than the open channel or pipe flow, here we use a Taylor-Couette (TC) facility \cite{gro16}, which is a closed system obeying global balances and at the same time allows for both accurate global and local measurements. The overall torque $\tau$ in TC flow to keep the cylinders at constant angular velocity, is connected with the spatially and time averaged energy dissipation rate $\epsilon_u$. 
This can be expressed in terms of the friction factor \cite{pop00,sch00,gro16}
 \begin{eqnarray}
c_f = {\tau\over  \ell \rho_f \nu^2 (\mathrm{Re}_i-\eta \mathrm{Re}_o)^2}= {\pi \eta \over (1-\eta)}    {\epsilon_u\over (U_i-\eta U_o)^3/(r_i+r_o) } . 
\label{tau-eps}
 \end{eqnarray}
Here
$U_{i,o}$ are the velocities of the inner resp. outer 
cylinder, $r_{i,o}$  their radii, $\nu$ the kinematic viscosity (together defining the
inner and outer Reynolds numbers $\mathrm{Re}_{i,o} = U_{i,o} d/\nu$), $\rho_f$ the density of the fluid, $\ell$ the height of the TC cell,
$d=r_o-r_i$ the gap width, 
and $\eta =  r_i/r_o$ the ratio between outer and inner cylinder radius. 
The key question now is: how does the friction factor $c_f$ depend on the (driving) Reynolds number $\mathrm{Re}_{i,o}$ and
how does wall roughness affect this relation?

Alternatively, the Reynolds number dependence of the friction factor $c_f$ can be expressed as a ``Nusselt number''
$\mathrm{Nu}_\omega =  \tau / (2\pi \ell \rho_f  J_{lam}^\omega )$ (i.e. the dimensionless angular velocity flux with the laminar flux 
$J^\omega_{lam} = 2\nu r_i^2 r_o^2
(\omega_i - \omega_o)/ (r_o^2 - r_i^2)$ \cite{eck07b})
depending on 
the Taylor number 
$\Ta= \frac{1}{64}\frac{(1+\eta)^4}{\eta^2} d^2 (r_i+r_o )^2(\omega_i - \omega_o)^2\nu^{-2}$ \cite{gro16}, 
with $\omega_{i,o}$ the angular velocity of the inner resp.\ outer cylinder. 
This notation $\mathrm{Nu}_\omega (\mathrm{Ta})$ stresses the analogy between TC flow and Rayleigh-B\'enard flow (RB) \cite{ahl09,loh10},
the flow in a box heated from below and cooled from above, 
where the Nusselt number $\mathrm{Nu}$ (the dimensionless heat flux) depends  on the Rayleigh number $\textrm{Ra}$ (the dimensionless
temperature difference). For that system Kraichnan \cite{kra62} had postulated a so-called ``ultimate scaling regime'' \cite{cha97,he12,he12a,hui12,ost14pd,gro16} 
 \begin{eqnarray}
\mathrm{Nu} \sim \mathrm{Ra}^{1/2} (\log \mathrm{Ra} )^{-3/2} 
\label{kra-rb1}
 \end{eqnarray}
(for fixed Prandtl number). 
 In analogy, such an ultimate regime also exists for TC flow,
 namely 
 \begin{eqnarray}
\mathrm{Nu}_\omega \sim \mathrm{Ta}^{1/2} (\log \mathrm{Ta} )^{-3/2}, 
\label{kra-tc1}
 \end{eqnarray}
 as worked out in Ref.\ 
 \cite{gro11}. 
 In fact, in that reference slightly different log-dependences were derived, namely 
 \begin{eqnarray} \label{kra-rb2}
 \mathrm{Nu} & \sim &  \mathrm{Ra}^{1/2} {\cal{L}}(\mathrm{Re}),  \quad \hbox{and}\\
 \mathrm{Nu}_\omega & \sim &  \mathrm{Ta}^{1/2} {\cal{L}}(\mathrm{Re}),
 \label{kra-tc2} 
 \end{eqnarray}
 where ${\cal{L}} (\mathrm{Re} (\mathrm{Ra}))$ resp.
 ${\cal{L}} (\mathrm{Re} (\mathrm{Ta}))$ 
 are logarithmic dependences (see Methods and also Ref.\  \cite{gro11}).  
 Irrespective of whether one takes the logarithmic dependences 
 (\ref{kra-rb1}) resp.\ 
 (\ref{kra-tc1}) or 
   (\ref{kra-rb2}) resp.\ 
 (\ref{kra-tc2}), for smooth walls 
 due to these log-corrections 
 the {\it effective} scaling exponent for the largest experimentally achievable 
 Rayleigh (Taylor) numbers is only around 0.38 and not 1/2, i.e., $\mathrm{Nu}\sim \mathrm{Ra}^{0.38}$ resp. $\mathrm{Nu}_\omega\sim \mathrm{Ta}^{0.38}$. 
 This effective exponent 0.38 has indeed been observed in large $\mathrm{Ra}$ RB experiments \cite{he12,he12a}, large $\mathrm{Ta}$ TC experiments \cite{hui12,gro16} and numerical simulations \cite{ost14pd,gro16}. 
The log-corrections, which are intimately connected with the logarithmic boundary layers \cite{ost16jfm}, thus
prevent the observation of the asymptotic ultimate regime exponent $1/2$, which is the exponent of mathematically strict upper bounds for RB and TC turbulence \cite{doe96,nic97,pla03}. Historically, whether such asymptotic 1/2 scaling exists or not has triggered enormous debate, see e.g. \cite{ahl09}. In the last two decades, great efforts have been put into reaching this regime with smooth boundaries, both experimentally and numerically. Today, this issue is often considered as one of the most important open problems in the
thermal convection community. In fact, the exponent $1/2$ has only been achieved with rough walls \cite{top17} as presumably a transient, local effective scaling, which saturates back to smaller exponent at even larger $\mathrm{Ra}$ \cite{ahl09,xie17,zhu17b}, or in
artificial configurations, such as numerical simulations of so-called ``homogeneous convective turbulence'' \cite{loh03} with periodic
boundary conditions and no boundary layers, or experimental realisations thereof such as in Refs.\ \cite{gib06,cho09}.

The asymptotic exponent $1/2$ in the $\mathrm{Nu}$ vs. $\mathrm{Ra}$ resp. $\mathrm{Nu}_\omega$ vs. $\mathrm{Ta}$ scaling law corresponds to a friction factor $c_f$ being {\it independent} of the Reynolds number. Vice versa, expressed in terms of the friction factor, Eqs.\ (\ref{kra-rb1}) to (\ref{kra-tc2}) can be written with a logarithmic dependence, analogous to the so-called Prandtl-von K\'arm\'an skin friction law \cite{kar21,sch00,pop00} for pipe flow, i.e.
 \begin{eqnarray}
1/\sqrt c_f = a \mathrm{log}_{10} (\Re_i \sqrt c_f) + b, 
\end{eqnarray}
which can be obtained by assuming that the boundary layer 
  profiles at each cylinder wall are 
  logarithmic and match at the middle gap  \cite{lat92,hui13,ost14pof}. Here $a$ and $b$ are fitting constants
  connected with the von K\'arm\'an constant $\kappa$.

 How to get rid of the log-correction and to thus achieve asymptotic ultimate turbulence with a 1/2 power law 
 or equivalently a Reynolds number independent friction factor for TC flow? The path we will follow here is to introduce wall roughness \cite{roc01,she96,ber03}. By combining direct numerical simulations (DNS) and experiments (EXP), we explore five decades of $\mathrm{Ta}$ to present conclusive evidence that the 1/2 power law can be realized, thus achieving the asymptotic ultimate regime. Moreover, we will give a theoretical justification for the
 findings based on measurements of the global and local flow structures and extend the analysis also to outer cylinder rotation.

  \begin{figure}[!h]
\begin{center}
\includegraphics[height=4.4in]{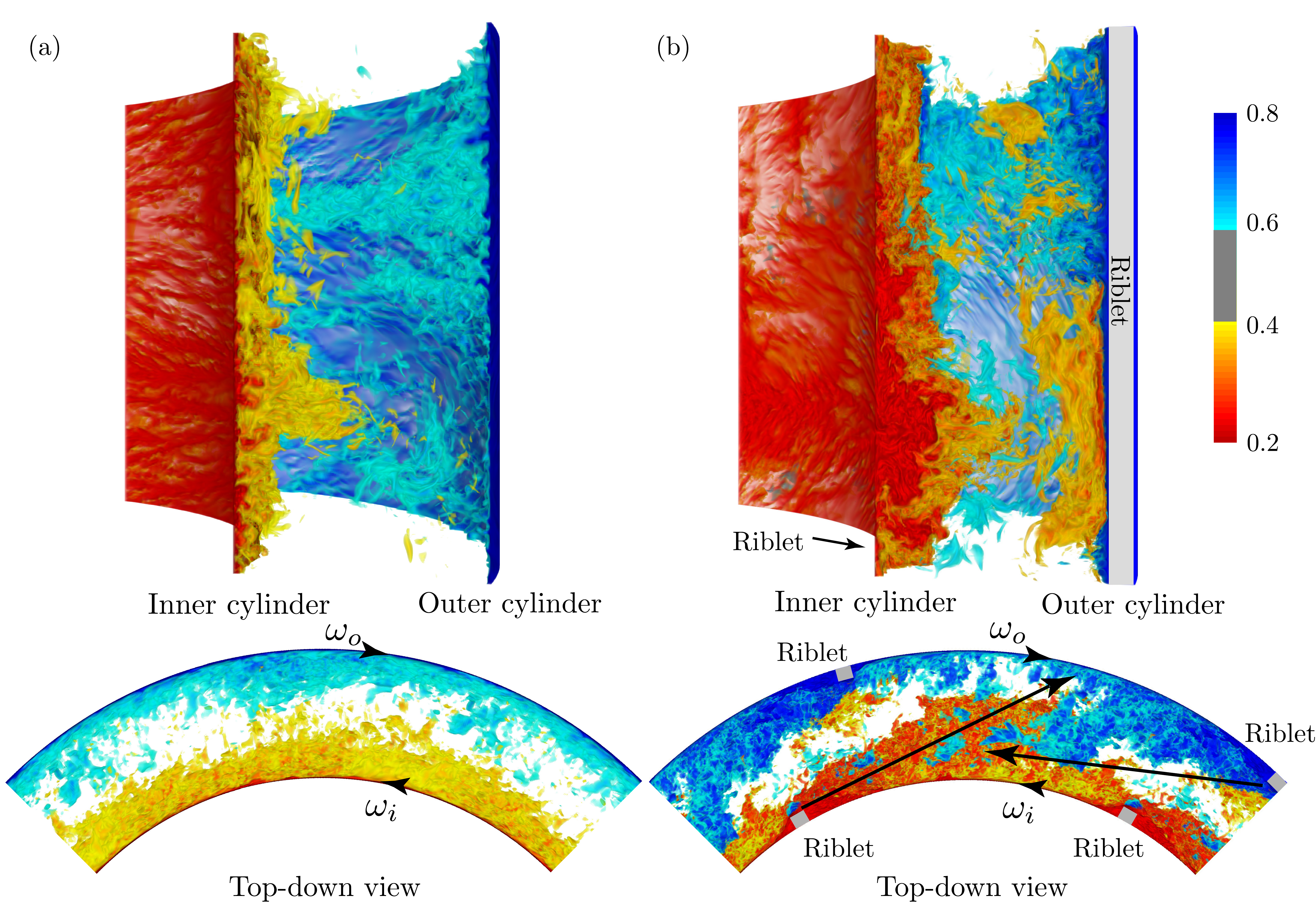}
\caption{
{\bf Plume structures for smooth and rough Taylor-Couette turbulence,} i.e. turbulent flow between two co-axial rotating cylinders, with inner cylinder rotates at angular velocity $\omega_i$ and outer cylinder at $\omega_o$. Here the volume renderings of azimuthal velocity at $\Ta=2.15\times10^9$ and Rossby number $\Ro^{-1}=-0.2$ are shown, from numerical simulations (see Methods for more details). {\bf a,} Both cylinders are smooth. The plumes are generated on both cylinders and form the structure of Taylor rolls and they are concentrated in local regions and can not reach the other cylinder. {\bf b,} Both cylinders are rough with 6 ribs of height $h=0.1d$. Even in the rough case, Taylor rolls still exist. Now the plumes are also generated on top of the roughness elements and shed to the opposing cylinder. The arrows in the top-down views illustrate the directions of plumes shedding. All plots share the same colormap, based on the value of the local angular velocity.}
\label{fig:fig1}
\end{center}
\end{figure}

Four cases will be considered: SS, SR, RS, and RR, where the first (second) letter specifies the configuration of the inner (outer) cylinder, which can be either rough (R) or smooth (S). In both DNS and EXP, the radius ratio between the two cylinders is $\eta=0.716$. The cylinders were made rough by attaching 1 to 192 vertical ribs with identical heights ranging from 1.5\% to 10\% of the gap width $d$ and a square cross-section over the entire TC cell on none, both, or either one of the cylinders (see Methods section). To give the reader an impression of the flow organization, 
 typical flow structures of a smooth case and a rough case are shown in Figs.\ \ref{fig:fig1}a, b, respectively.

\section{Global scaling relations} 

\begin{figure}[htbp]
\begin{center}
\includegraphics[width=4.5in]{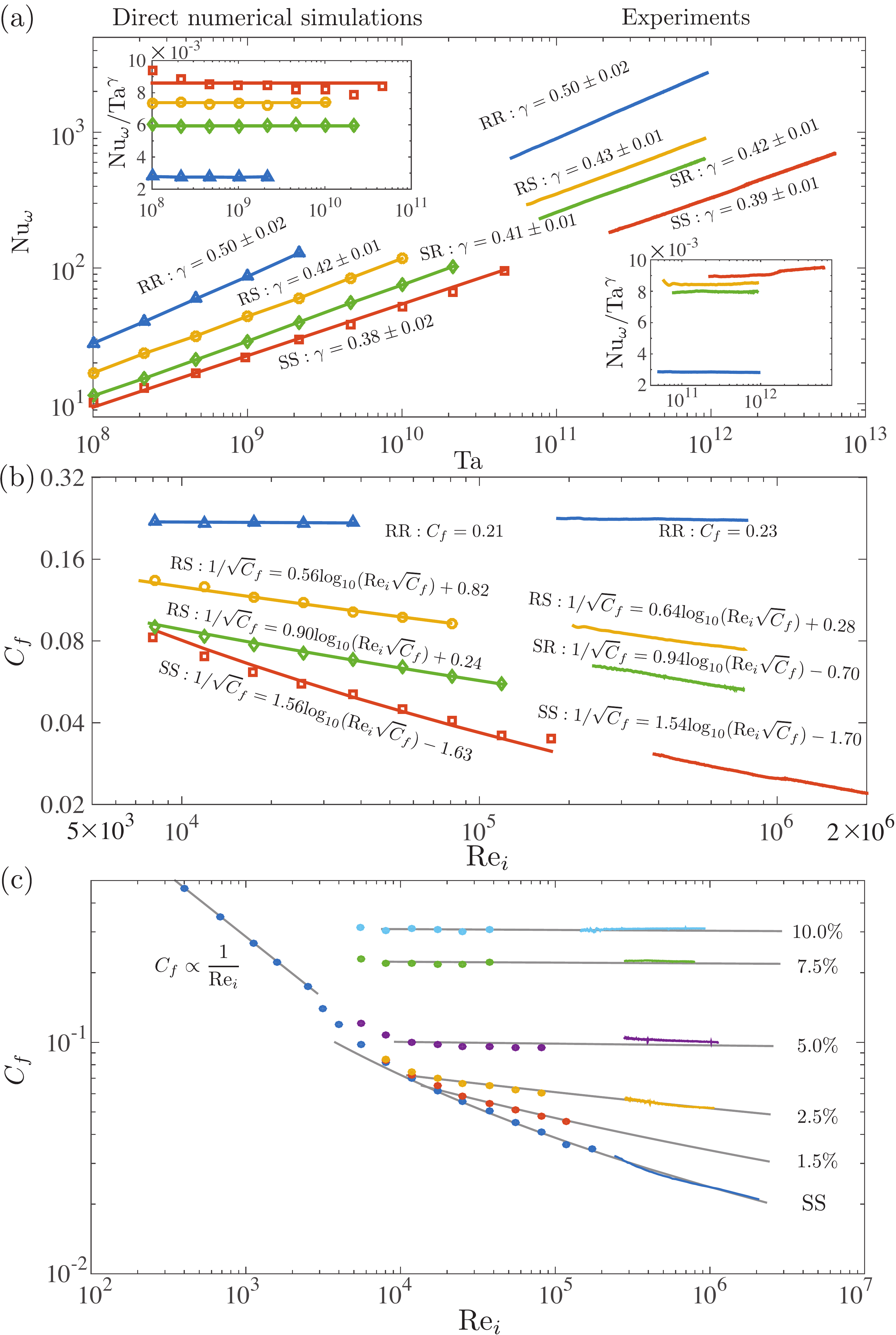}
\caption{ 
{\bf Global torque and friction factor scalings in both DNS (symbols), and experiments (colored lines).} {\bf a,} The dimensionless torque as a function of Taylor number $\Ta$. Four cases are shown: (SS) both cylinders smooth; (SR) smooth inner, rough outer; (RS) rough inner, smooth outer; and (RR) both cylinders rough, with the exponent $\gamma$ in the power law relation $\textrm{Nu}_\omega \sim \mathrm{Ta}^\gamma$ shown for every case. The insets depict the compensated plots $\Nu/\Ta^\gamma$, showing the quality of the scaling. {\bf b,} The friction factor $c_f$ as a function of the inner cylinder Reynolds number $\Re_i$. The lines show the best fits of the Prandtl friction law $1/\sqrt c_f=a \mathrm{log} _{10}(\Re_i \sqrt c_f)+b$, with all prefactors shown in the figures. For {\bf a} and {\bf b}, 6 ribs were used and the roughness height is $h=0.075d$. For the RR case, $\Re_i$ independent friction factors are revealed. {\bf c,} The friction factor $c_f$ for RR cases with 6 ribs of different heights, ranging from 1.5\% to 10\% of the gap width $d$.
}
\label{fig:fig2}
\end{center}
\end{figure}

In this section we address the question of how roughness modifies the global scaling relations. First, we focus on the cases of 6 ribs with identical heights $h=0.075d$, both numerically and experimentally. The global dimensionless torques, $\Nu \sim \Ta^{\gamma}$, for the four cases, with increasing $\Ta$ and fixed outer cylinder, are shown in Fig.\ \ref{fig:fig2}a. Combining EXPs and DNSs, the range of Taylor number studied here extends over five decades. Similarly to what was shown in various recent studies \cite{he12,gil11,ost14pof,ost14pd,bra13,gro16}, for the SS case, an effective scaling of  $\Nu \sim \Ta^{0.38\pm0.02}$ is observed in the DNS, corresponding to the ultimate regime with logarithmic corrections \cite{kra62,gro11}. A very similar scaling exponent $\Nu \sim \Ta^{0.39 \pm 0.01}$ is found in EXPs, demonstrating the excellent agreement between DNS and EXPs. 

Dramatic enhancements of the torques are clearly observed with the introduction of wall roughness, resulting in the transition of $\Nu$ from $\mathcal{O}(10^2)$ to $\mathcal{O}(10^3)$. Specifically, when only a single cylinder is rough, the logarithmic corrections reduce and the scaling exponents marginally increase, implying that the scaling is dominated by the single smooth wall. For the RR case, the best power law fits give $\Nu \sim \Ta^{0.50\pm0.02}$, both for the numerical and experimental data, suggesting that the logarithmic corrections are thoroughly canceled. This state with the scaling exponent $1/2$ corresponds to the {\it asymptotic ultimate turbulence} predicted by Kraichnan \cite{kra62}. The compensated plots of insets of $\Nu/\Ta^{\gamma}$ show the robustness and the quality of the scalings. 

When expressing the relation between the global transport properties and the driving force in terms of
 the Reynolds number dependence of the friction factor $c_f$, we obtain  Fig.~\ref{fig:fig2}b. 
 For the SS case, the fitting parameters $a$ and $b$ yield a von K\'arm\'an constant $\kappa=0.44\pm0.01$, which is slightly larger than the standard value of 0.41 due to the curvature effect \cite{hui13,ost16jfm,gro14}. This agrees very well with the previous measurements on TC with smooth walls \cite{lew99}. For the RR case, in both DNS and EXP, for large enough driving the friction factor $c_f$ is found to be independent of $\Re_i$, but dependent on roughness height, namely $c_f=0.21$ in the DNS and $c_f=0.23$ in the EXP for roughness height $h=0.075d$, thus showing good agreement also for the rough cases. The results here are consistent with the asymptotic ultimate regime scaling $1/2$ for $\Nu$ and indicate that the Prandtl-von K\'arm\'an log-law of the wall \cite{sch00,pop00} with wall roughness can be independent of $\Re_i$ \cite{nik33,sch00,pop00,jim04,fla14}, which has been verified recently for Taylor-Couette flow \cite{zhu17}. For the RS and SR cases, one boundary is rough and the other is smooth such that the friction law lies in between RR and SS lines. 
 
We further show the RR case with ribs of different heights, ranging from 1.5\% to 10\% of the gap width $d$ in Fig.\ \ref{fig:fig2}c, displaying its similarity with the Nikuradse \cite{nik33} and Moody \cite{moo44} diagrams for pipe flow. It can be seen that once $h\geqslant 0.05d$ and $\textrm{Re}_i \geqslant 8.1\times10^3$ ($\Ta \geqslant 10^8$), the asymptotic ultimate regime can always be achieved in both DNS and EXP.

 Analogously, we note that in pipe flow, the same phenomenon of Reynolds number independent friction factor with wall roughness was observed in the fully rough regime \cite{nik33,sch00,pop00,jim04,fla14}, where the characteristic heights of the roughness elements in wall units $h^+ >70$ \cite{sch00,pop00}. In contrast, for $\Ta=10^8$, for the roughness height $h/d=0.05$, $h/d=0.075$, and $h/d=0.10$, $h^+=51$, $h^+=61$, and $h^+=71$, respectively. Indeed, almost all of our data are in the fully rough regime for cases with $h\geqslant 0.05d$ and $\Ta \geqslant 10^8$, thus corroborating the current conclusion that adopting wall roughness is one way to achieve asymptotic ultimate turbulence in TC.

We now interpret the asymptotic ultimate torque scalings through an extension of the Grossmann-Lohse (GL) theory \cite{gro11}, by accounting for the Prandtl-von K\'arm\'an log-law of the wall \cite{sch00} in the presence of roughness. To demonstrate this extension, for simplicity we take as example the case of only inner cylinder rotation. For a smooth wall, the energy dissipation rate in the log region scales with $\epsilon_{u} d^4/\nu^3 \sim \Re_i^3(u_\tau/U)^3 \mathrm{ln} (\Re_i u_\tau/U)$ \cite{gro11}, which stems from the integration of the Prandtl-von K\'arm\'an log-law of the wall, where $u_\tau$ is the friction velocity and $U$ the velocity of the inner cylinder. The log term in the law is dependent on $\Re_i$, which is the origin of the logarithmic correction term ${\cal{L}}(\Re)=(u_\tau/U)^3 \mathrm{ln} (\Re_i u_\tau/U)$ and thus for the deviation from the asymptotic ultimate regime scaling $\epsilon_ud^4/\nu^3 \sim \Re_i^3$, leading to a decrease of the effective scaling exponent. However, with roughness, as stated before, the log term in the law of the wall becomes independent of $\Re_i$  \cite{sch00,pop00,jim04,fla14,zhu17}, which correspondingly renders this correction \textit{constant}. Translating this argument for the energy dissipation rate $\epsilon_{u} (\mathrm{Re_i})$ back to the dimensionless torque $\Nu$ and the driving force $\Ta$ \cite{eck07b}, we obtain $\Nu \sim \Ta^{1/2}$, i.e. the effect of the logarithmic term on the scaling vanishes; see Methods for details.

\begin{figure}[!htp]
\begin{center}
   \includegraphics[width=3.2in]{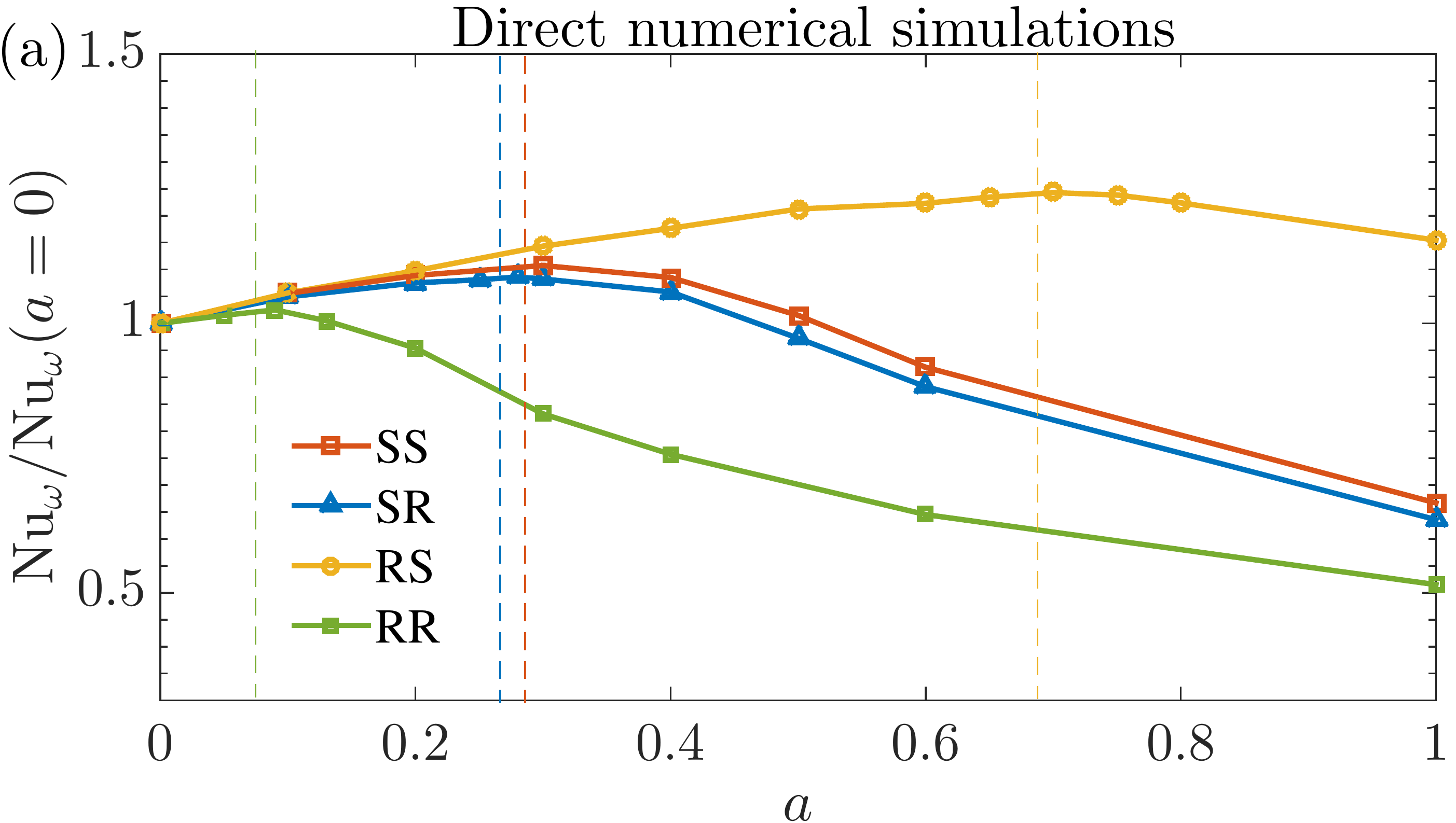}
    \includegraphics[width=3.2in]{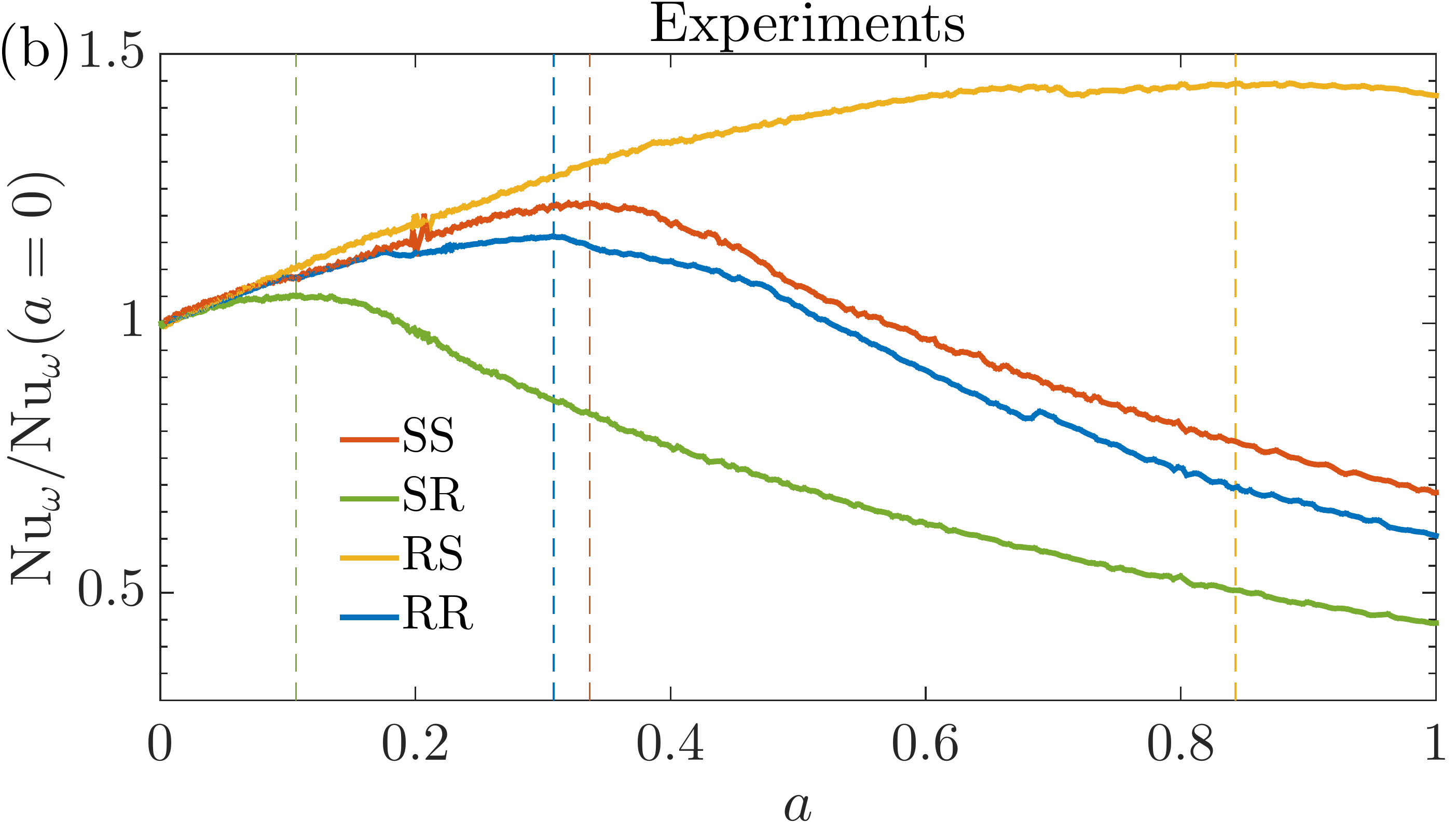}
\caption{{\bf Optimal transport peak.} $\Nu$ as function of $a$ for constant driving strength, normalized by its value for $a=0$. For both EXP and DNS, 6 ribs were used and the roughness height is $h=0.075d$. {\bf a,} DNSs with $\Ta=1\times 10^9$. The optimal transport peaks are located at  $a_{opt,SS}^{DNS}=0.30 \pm0.03$, $a_{opt,SR}^{DNS}=0.09\pm0.03$, $a_{opt,RS}^{DNS}=0.69\pm0.05$ and $a_{opt,RR}^{DNS}=0.28\pm0.03$. {\bf b,}  Experiments with $\Ta=4\times 10^{11} $. The optimal transport peaks for the four cases are located at $a_{opt,SS}^{EXP}=0.34$, $a_{opt,SR}^{EXP}=0.11$, $a_{opt,RS}^{EXP}=0.84$ and $a_{opt,RR}^{EXP}=0.31$. All optimal transport peaks are indicated by the dashed lines, with the respective colors. }
\label{fig:Nu_a}
\end{center}
\end{figure}

One distinct difference between TC and pipe flow is that in a TC system, the inner and outer cylinders can rotate independently, resulting in a second control parameter, namely the rotation ratio $a=-\omega_o/\omega_i$ of the two cylinders. Just as for smooth walls \cite{gil11,gro16}, 
also for rough walls the $\Nu \sim \Ta^{\gamma}$ scaling exponents are independent of the rotation ratio $a$ in the studied rotation ratio regime; see Extended Data Fig.\ \ref{fig:fig8}. As known since Taylor \cite{tay23b}, the inner cylinder rotation has a destabilizing effect on the flow, whereas outer cylinder rotation has a stabilizing effect. For TC flow with smooth walls, it was found that the optimal transport rotation ratio $a_{opt}$ between the two cylinders, where the torque reaches the maximum for a specific driving $\Ta$, is around $a_{opt}= 0.36$ \cite{,bra13b,hui14}, and not zero, as one may have assumed. This is attributed to the existence of the strong Taylor rolls between the counter-rotating cylinders when $a \approx a_{opt}$. Only for strong enough counter-rotation ($a>a_{opt}$) does the stabilization through the counter-rotating outer cylinder take over \cite{gil12}. Here, we address the question whether this optimal transport rotation ratio shifts or stays the same in the presence of roughness. The results are shown in Fig.\ \ref{fig:Nu_a}. 
We find that when either one of the cylinders is rough, the effect of that rough cylinder is enhanced in several ways, as we will now elaborate.

In the SR case, $a_{opt,{SR}}^{DNS}=0.09 \pm 0.03$ and  $a_{opt,{SR}}^{EXP}=0.11$, i.e.\ little outer cylinder rotation is necessary to reduce the angular velocity transport with the help of the roughness elements on it, which are thus not so effective. In contrast, a rough {\it inner} cylinder is much more effective to enhance the momentum transport. The optimal transport peak for the RS case occurs at much larger rotation ratio, $a_{opt,{RS}}^{DNS}=0.69 \pm 0.05$ and  $a_{opt,{RS}}^{EXP}=0.84$, as very strong outer cylinder rotation is needed to suppress turbulence originating from the rough inner cylinder. In this case the stabilizing effect of the smooth outer cylinder becomes inefficient. 

Finally, in the RR case, the effects of the inner cylinder and outer cylinder are balanced in a similar way as in the SS case, resulting in similar values of $a_{opt,{RR}}^{DNS}=0.28  \pm 0.03$ and  $a_{opt,{RR}}^{EXP}=0.31$ as found in the SS case ($a_{opt,{SS}}^{DNS}=0.30  \pm 0.03 $ and  $a_{opt,{SS}}^{EXP}=0.34$). At optimal rotation ratio $a_{opt}$, the enhanced shear is caused by Taylor rolls \cite{hui14,ost14pd,cho14,mar14jfm}. This indicates that even in the presence of roughness, Taylor rolls still exist, as visible in Fig.\ \ref{fig:fig1}b. We further notice that the optimal transport properties are dependent on the roughness height, as shown in Extended Data Fig.\ \ref{fig:fig9}. As expected, when the roughness height is smaller, $a_{opt}$ for SR and RS cases are closer to $a_{opt}$ for the SS case. On the contrary, when the roughness height is larger, $a_{opt}$ for SR and RS cases deviates more from $a_{opt}$ for the SS case. This can be clearly seen from Extended Data Fig.\ \ref{fig:fig10}.

\begin{figure}[htbp]
     \includegraphics[width=3.2in]{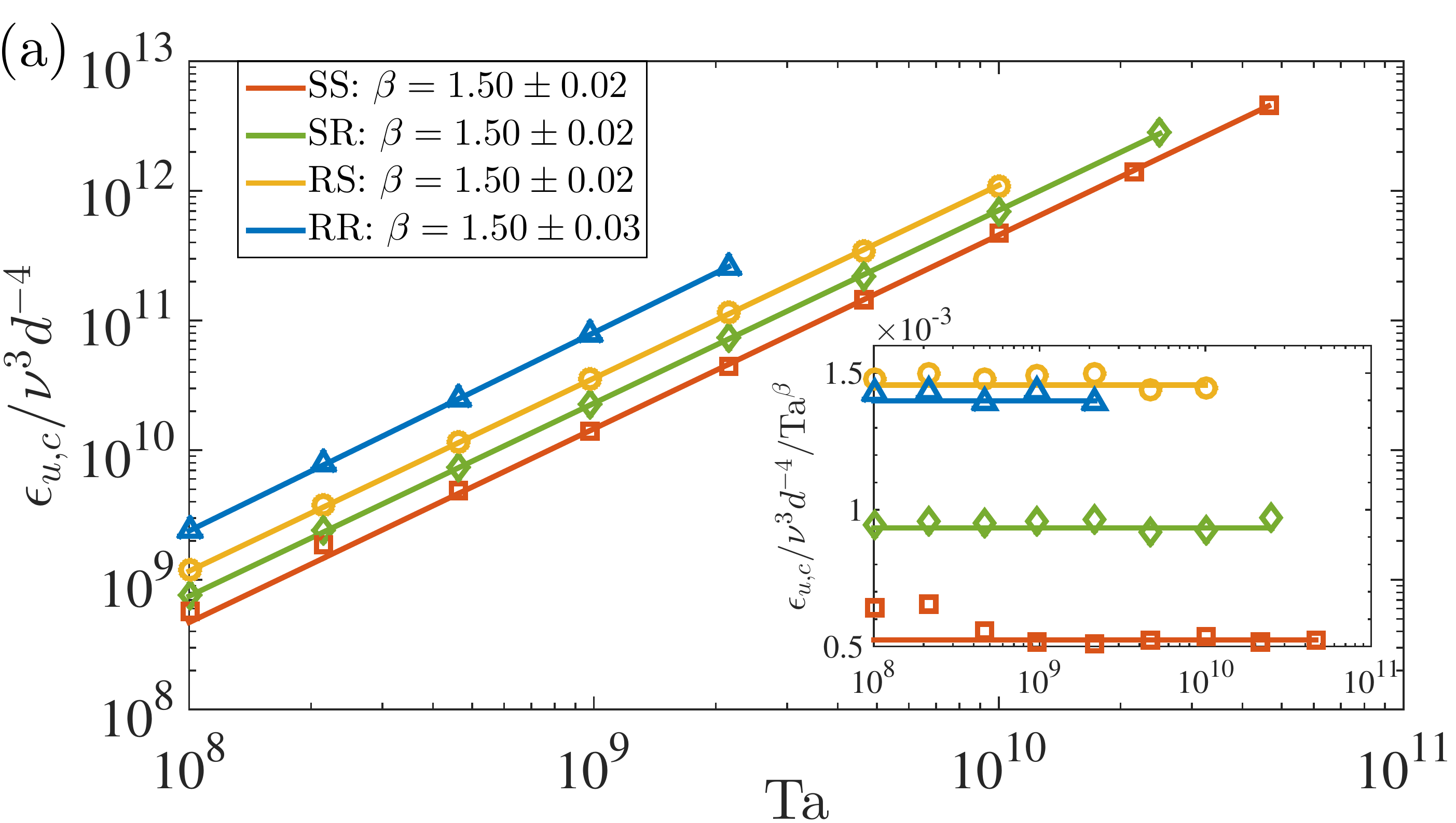}
    \includegraphics[width=3.2in]{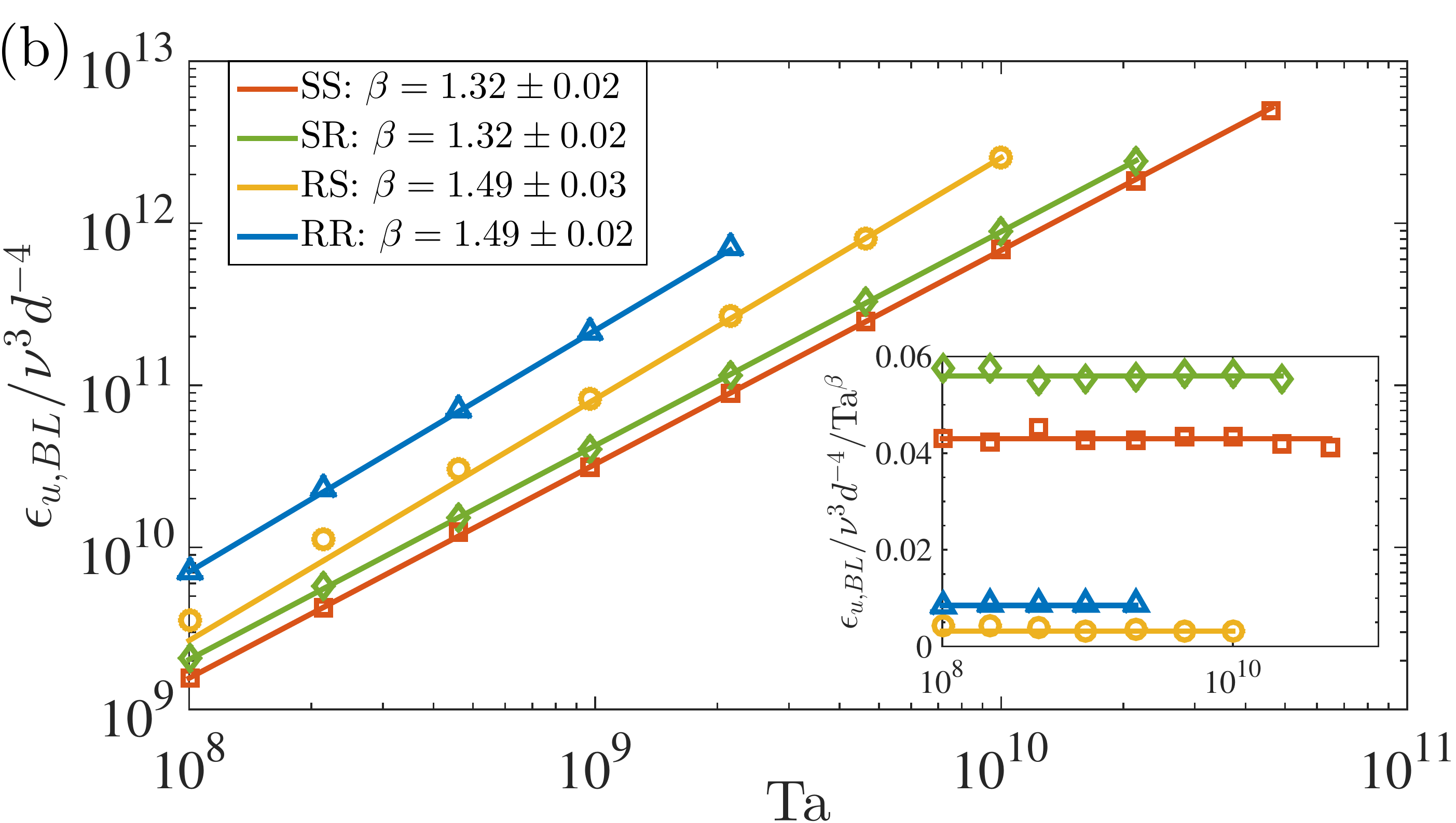}
\caption{{\bf Local energy dissipation rate from simulations.} Local energy dissipation rate in the bulk $\epsilon_{u,c}$ (at the center of the gap, averaged over the height) and in the inner cylinder boundary layer $\epsilon_{u,BL}$ (averaged in the range from the wall to the distance corresponding to the maximum root mean square of the azimuthal velocity) as a function of $\Ta$. For the rough cases, 6 ribs were used and the roughness height is $h=0.1d$. The symbols are the numerical data and the lines show the best fits.\ {\bf a,} The bulk energy dissipation rate follows $\epsilon_{u,c} \sim \Ta^{1.50} \sim Re_i^3$, irrespective of whether the wall is smooth or rough.\ {\bf b,} The boundary layer dissipation rate at the inner wall follows $\epsilon_{u,BL} \sim \Ta^{1.32}$ for the cases with smooth walls, while it scales with $\epsilon_{u,BL} \sim \Ta^{1.50}$ for the cases with rough inner wall.}
\label{fig4}
\end{figure}

\section{Local flow organization and profiles} 

Till now, we have focused on the global transport properties. However, the details of the boundary layer-bulk interaction, and in particular how the local scalings of the energy dissipation rates affect the global ones, are still unknown. To verify above sketched theory, from our DNS data we split the mean energy dissipation rate (Eq. \ref{tau-eps}) into boundary layer and bulk contributions, following the GL approach \cite{gro00,gro01}. In Fig.~\ref{fig4}(a), the local energy dissipation rates at mid-gap $\epsilon_{u,c}$  are shown as a function of $\Ta$ (only inner cylinder rotation, $a=0$). It is clear that no matter whether the wall is smooth or rough, the bulk energy dissipation rate follows $\epsilon_{u,c}\sim \Ta^{3/2} \sim 
\Re_i^3$, which corresponds to the asymptotic ultimate regime without any logarithmic correction. In analogy, for RB turbulence, the same scaling exponent $\epsilon_{u,c}\sim \Ra^{3/2}$ was reported in Refs.\ \cite{sha08,ni11b}. Therefore, the crucial element determining the overall scaling is the dissipation rate in the boundary layer. To further confirm this, in Fig.~\ref{fig4}(b) we show the local energy dissipation rates of the boundary layer $\epsilon_{u,BL}$ (averaged in the range from the wall to the distance corresponding to the maximum root mean square of the azimuthal velocity). For the case with smooth walls, we find $\epsilon_{u,BL} \sim \Ta^{1.32}$ because of the $\Re_i$-dependent velocity profile, while for the boundary layers at rough walls we have $\epsilon_{u,BL} \sim \Ta^{3/2}$ because, as shown above, roughness cancels out the $\Re_i$-dependence in ${\cal L}(\Re_i)$ and thus restores the asymptotic ultimate regime scaling. 
The competition between the boundary layer and bulk
ultimately determines the global scalings.

\begin{figure}[!htp]
\includegraphics[width=3.2in]{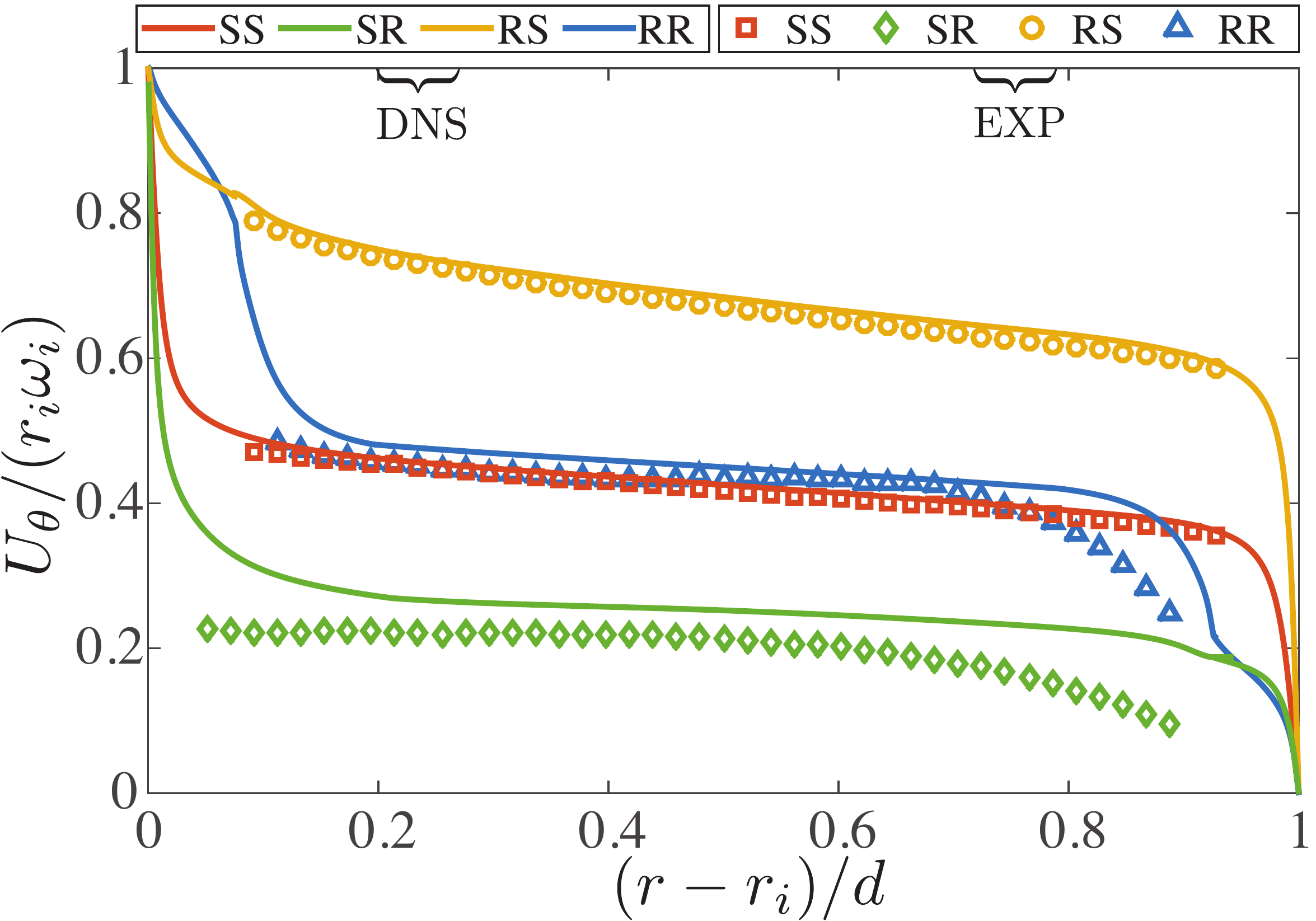}
\caption{{\bf Mean velocity profiles.} Normalized azimuthal velocity $U_{\theta}(r)/(r_i \omega_i)$ profiles as a function of the normalized radius $(r-r_i)/d$ for only inner cylinder rotation. For both EXP and DNS, 6 ribs were used and the roughness height is $h=0.075d$. Experimental and numerical data are shown in the same figure. EXP: $\Re_i = 5\times 10^5$ and DNS: $\Re_i = 3.74\times 10^4$. The experimental results were obtained using PIV. }
\label{fig:vel_prof}
\end{figure}

We now detail the origin of the enhanced torque. With roughness, the main contribution to the torque originates from the pressure differences between the side surfaces of rough elements, rather than from viscous forces \cite{sch00,pop00,jim04,fla14,zhu17}.  With roughness, we therefore expect the shear rate close to the rough wall to decrease significantly, as compared to the smooth case. This is clearly shown in Fig.\  \ref{fig:vel_prof}: with smooth cylinders, the normalized velocity profiles are characterized by a bulk region in which the velocity is relatively constant, $U_{\theta}=0.45r_i \omega_i$ (whereas for pipe flow, this is not the case, see Extended Data Fig.\ \ref{fig:fig12}). In case one single cylinder is rough, the bulk velocity is completely dominated by the velocity of the rough cylinder, or in other words, the bulk is enslaved to the rough wall. In the RR case, as there the torque is dominated by pressure forces, the shear rate at the rough cylinder is still smaller as compared to the smooth case. The implication is that with roughness, a larger fraction of energy dissipates in the bulk, and thus the system becomes bulk dominant. As mentioned before, the bulk energy dissipation rate follows $\epsilon_{u,c}\sim \Ta^{3/2}$, which implies the asymptotic ultimate regime. The more the bulk dominates the energy dissipation rate, the better the asymptotic ultimate regime manifests itself. This is indeed verified by the flow structure in Fig.\  \ref{fig:fig1}, where for the rough case, the plumes shedding from the roughness elements on one wall elongate towards the other wall and push more energetic fluid elements into the bulk, as compared to the smooth case, leading to more energy dissipation in the bulk.

\section{Controlling ultimate turbulence}
\begin{figure}[!h]
\begin{center}
\includegraphics[width=6.4in]{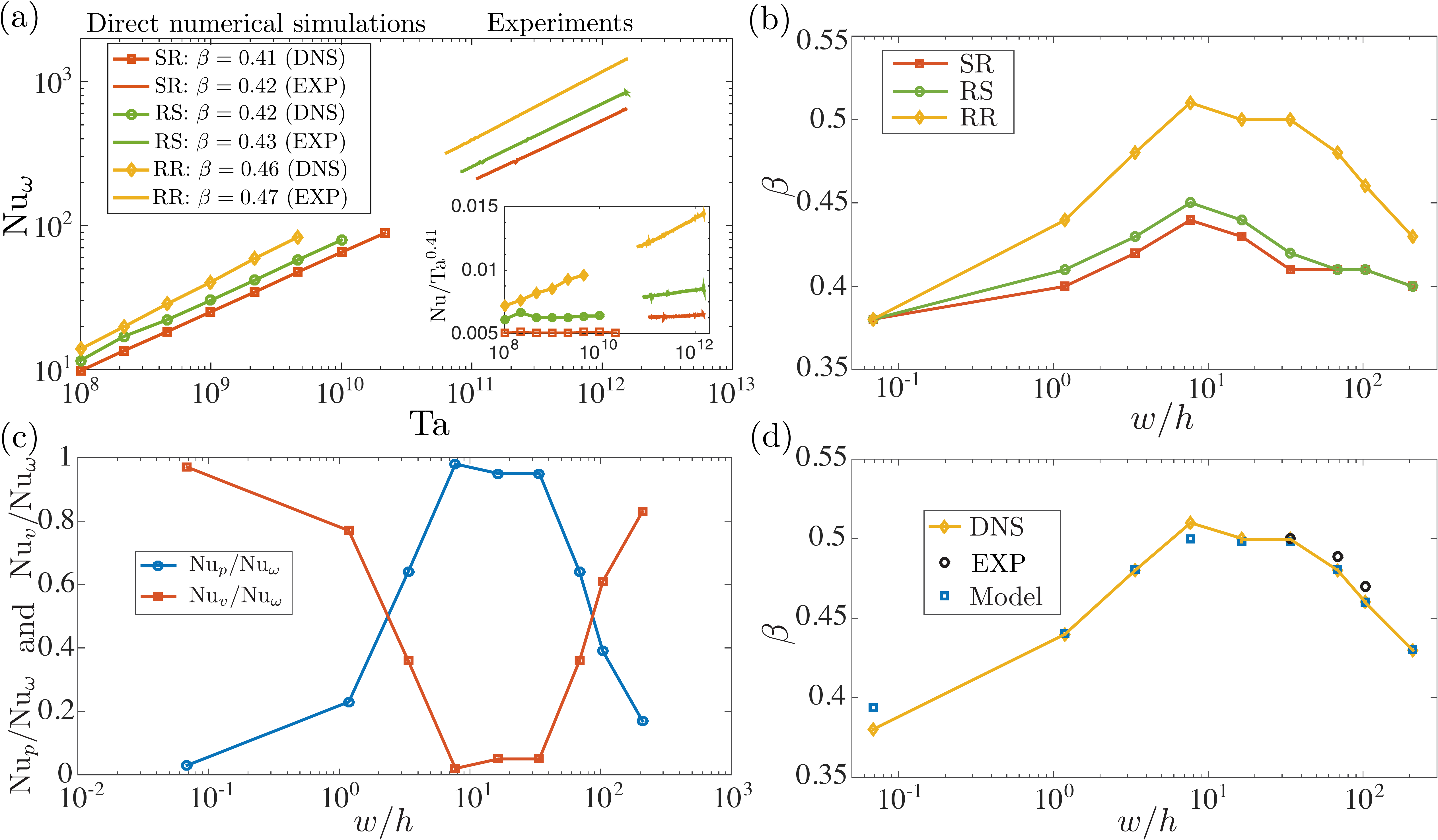}
\caption{ 
{\bf Dependence on the roughness density.} {\bf a,} The dimensionless torque as a function of Taylor number $\Ta$: DNS (left part), and experiments (right part) for the case of two ribs with height $h=0.075d$. For the RR case, the asymptotic ultimate regime is not yet achieved in this situation, in contrast to Fig.\ 2, when there are six ribs, for which the exponent is 0.5. {\bf b,} Effective scaling exponent $\beta$ for varying the gap width $w/h$ between the ribs, where $h$ is the height of the roughness. The number of ribs varies from 1 to 196 and correspondingly, the gap width $w/h$ varies from 208.44 to 0.07 at the inner cylinder. To get each $\beta$, five simulations between $\Ta=10^8$ and $\Ta=10^9$ were performed. {\bf c,} Contributions $\textrm{Nu}_p$ (from pressure drag) and $\textrm{Nu}_v$ (from viscous drag) to the global $\Nu$ at $\Ta=4.6\times10^8$ with varying the gap width $w/h$ between the ribs. The data are collected from DNS. The separation into the two parts is performed at the inner cylinder for the RR case. Clearly, when the pressure forces are dominant, $\beta$ is closer to 1/2 and when viscous forces are dominant, $\beta$ is closer to 0.38 (Fig.\ \ref{fig:fig6} (b)). {\bf d,} Comparison of the effective scaling exponent $\beta$ between the DNS results (RR case), EXP results (RR case), and the model results (based on Eq. \ref{eqn1}) with varying gap width $w/h$ between the ribs.
}
\label{fig:fig6}
\end{center}
\end{figure}

To bridge the gap between the effective ultimate scaling exponent 0.38 for the smooth case \cite{he12,he12a,ost14pd,gro16} and the asymptotic ultimate scaling exponent 0.5 for the RR case and thus to actively control ultimate turbulence, we vary the density of the roughness elements while keeping the height of the riblets fixed at 7.5\% of the gap width. To show how this will change the results, as an example, in Fig.\ \ref{fig:fig6}(a), we show the $\Nu$ vs. $\Ta$ scaling for the case of 2 ribs (very sparse). The effective scaling exponent $\gamma$ for the RR case is then smaller than $0.5$ (i.e. 0.47), so the asymptotic ultimate regime is not yet achieved in this situation, in contrast to Fig.\ 2, when there are six ribs, for which $\gamma=0.5$. We then continuously vary the number of ribs from 1 (very sparse) to 192 (very dense). Correspondingly, the spacing between the rough elements $w/h$ mounted on the inner wall varies from 208.44 to 0.07. We note that in pipe and BL flows, there is a distinction between k- and d-types of roughness, and a close spacing will make the roughness behave more like d-type roughness compared with k-type roughness \cite{jim04,fla14}. In Fig.\ \ref{fig:fig6}(b), we see that the effective scaling exponent is continuously changing with $w/h$. There is an optimal $w/h =7$ where the effective scaling exponent is the largest, corresponding to k-type roughness. To explain why the effective scaling exponent depends on $w/h$, in Fig.\ \ref{fig:fig6}(c) we split the global $\Nu$ into two parts, namely the viscous force contribution ($\textrm{Nu}_v$) and the pressure force contribution ($\textrm{Nu}_p$). Clearly, when the effective scaling exponent is higher, the pressure forces are more dominant.

We propose a simple model which can recover the effective scaling exponent. The model is based on the fact that in the smooth case, only viscous forces contribute to $\Nu$, resulting in $\Nu \sim \Ta^{0.38}$. In contrast, when the pressure forces take over, we have $\Nu \sim \Ta^{0.5}$. Therefore, in the spirit of GL theory of RB \cite{gro00}, we combine these contributions to set 
\begin{eqnarray}
\Nu=a\Ta^{0.38}+b\Ta^{0.5} \approx c\Ta^{\gamma_m}, 
\label{eqn1}
\end{eqnarray}
where $a=\textrm{Nu}_v/\Ta^{0.38}$ and $b=\textrm{Nu}_p/\Ta^{0.5}$ are the prefactors of the separated scalings for $\textrm{Nu}_v$ and $\textrm{Nu}_p$, respectively, which are roughness height dependent, and ${\gamma_m}$ is the effective local exponent predicted by the model. Here for the $h=0.075d$ case we use the separation shown in Fig.\ \ref{fig:fig6}(c) at $\Ta=4.6\times10^8$ to determine $a$, $b$, and hence the effective exponent ${\gamma_m}$ (other $\Ta$ can also be used and the results are similar). It can be seen that the model gives very good agreement with the DNS and EXP values (Fig.\ \ref{fig:fig6} (d)).

\section{Conclusions and outlook} 

The various wall roughness studies on turbulence in closed systems \cite{she96,du00,roc01,ber03,tis11,wei14} have resulted in quite different scaling exponents for the transport versus the driving forces, i.e. there has been no consensus \cite{ahl09} on whether the asymptotic ultimate turbulence 1/2 power law exists or not, a concept that was postulated 50 years ago by R.\ Kraichnan \cite{kra62}. Here, with both strong experimental and numerical evidence, we have demonstrated that the asymptotic ultimate regime scaling exponent 1/2, corresponding to the upper limit of transport, can be realized through the implementation of wall roughness in TC turbulence. We further showed that different number of roughness elements can tune the scaling exponents and optimal transport properties, thus paving the way to control ultimate turbulence. The insight gained from this study provides valuable guidance for any rotating and thermally driven turbulence with wall roughness in the ultimate regime, which is useful for a wide range of applications in industrial, geophysical, meteorological, and oceanographical flows.

\newpage

\newpage

\newpage
\section{Methods}


\subsection{Experimental methods}

\subsubsection{Experimental apparatus}
\begin{figure}[htp]
\begin{center}
\includegraphics[scale=1]{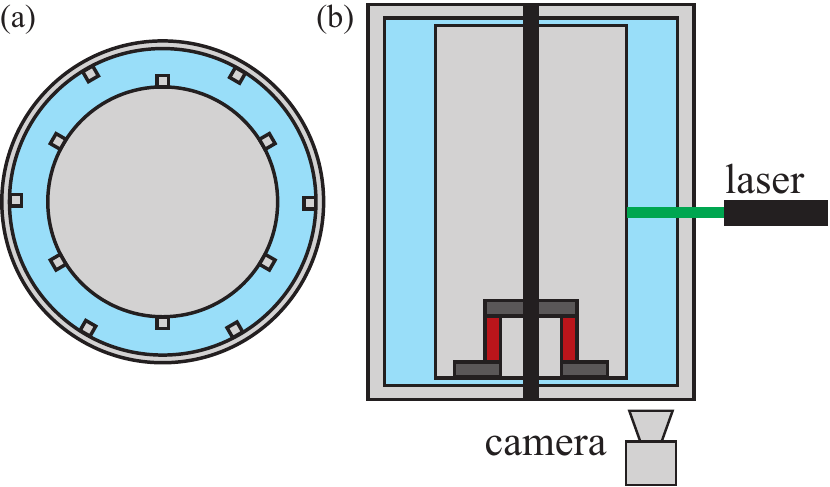}
\caption{{\bf Experimental setup} (a) Schematic of the top view of both the experimental and numerical setup for the (RR) case, i.e. both cylinders rough. The six ribs (not to scale), which are perpendicular to the axis of rotation, have a square cross section and extend over the entire height of the cylinders. Their size is 5\%, 7.5\%, and 10\% of the gap width for both the experiments and simulations. (b) Vertical cross-section of the experimental setup, showing the position of the torque sensor and the PIV setup. For the PIV measurements, the laser illuminates the horizontal ($r, \theta$) plane at mid-height, $z = L/2$, see the Methods section.}
\label{fig:setup}
\end{center}
\end{figure}

\begin{figure}[!h]
\begin{center}
\includegraphics[width=4.0in]{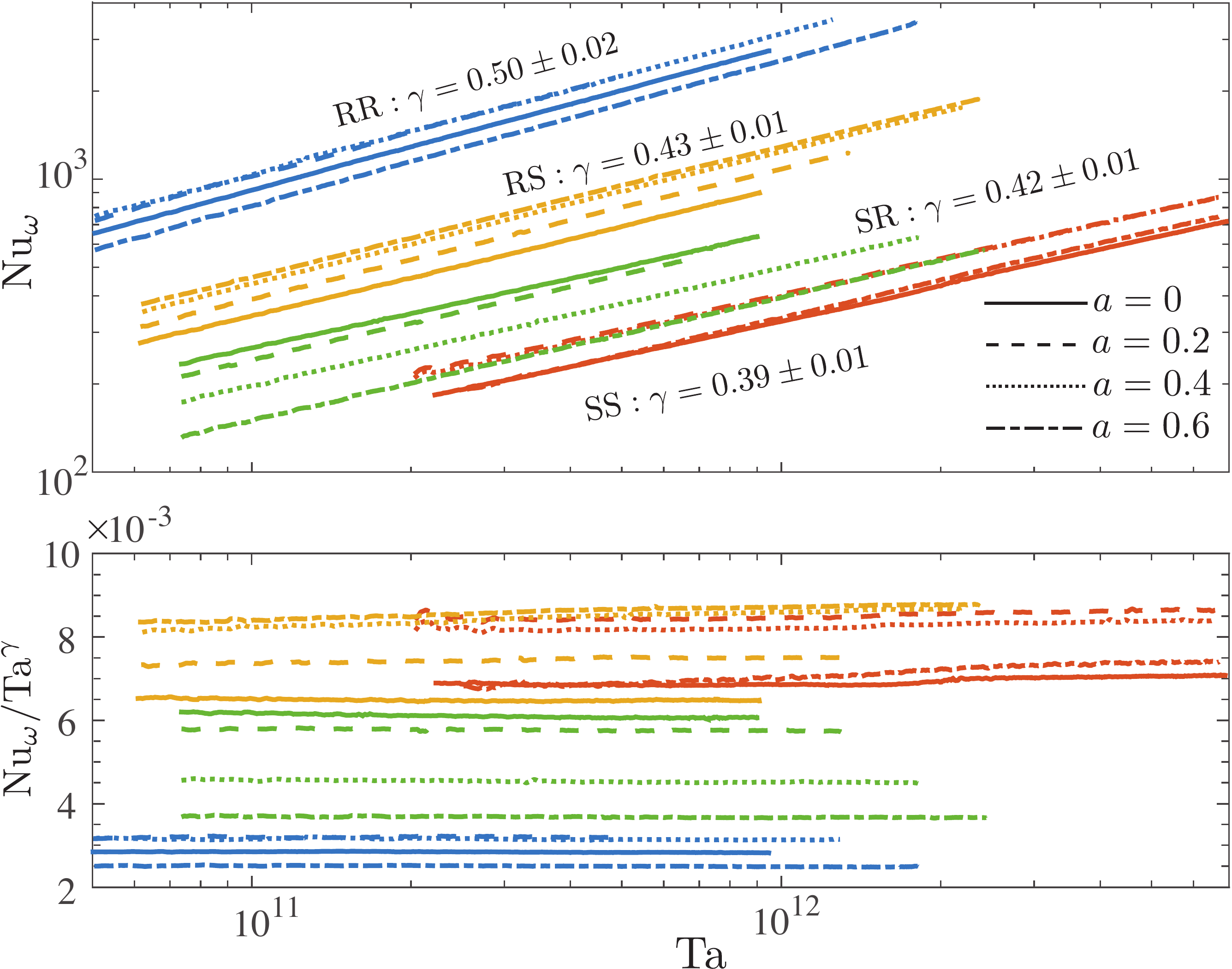}
\caption{ 
{\bf Torque scalings with different rotation ratio $a=-\omega_o/\omega_i$ from EXP.} Four cases are shown: (SS) both cylinders smooth; (SR) smooth inner, rough outer; (RS) rough inner, smooth outer; and (RR) both cylinders rough, with the exponent $\gamma$ in the power law relation $\textrm{Nu}_\omega \sim \mathrm{Ta}^\gamma$ shown for every case. The compensated plots $\Nu/\Ta^\gamma$ show the quality of the scaling. Clearly in the considered rotation ratio range, torque scalings are independent of the rotation ratio in all four cases.
}
\label{fig:fig8}
\end{center}
\end{figure}

\begin{figure}[!htp]
  \centering
 \begin{minipage}[b]{0.49\textwidth}
   \includegraphics[width=3.4in]{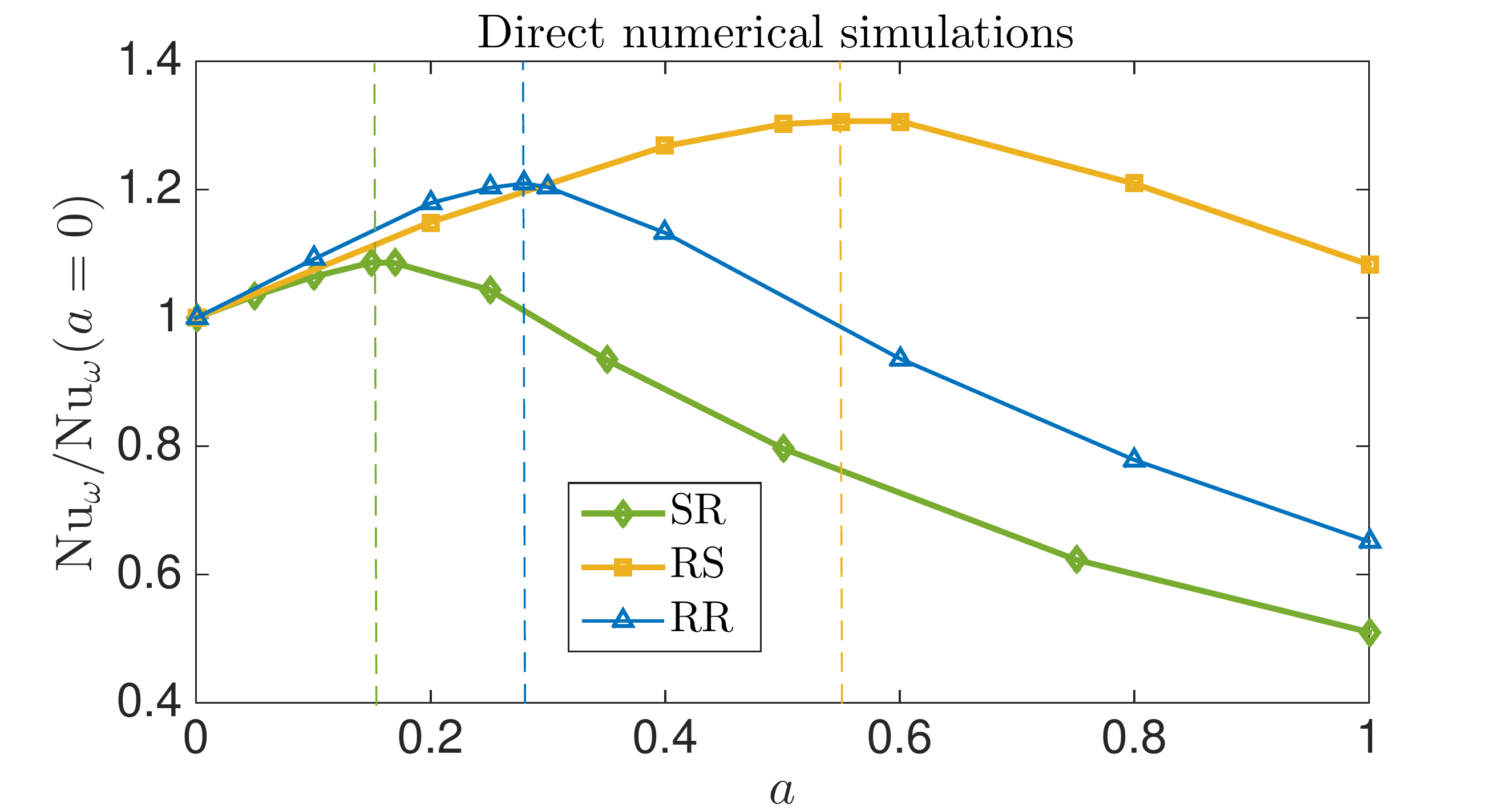}
 \end{minipage}
  \begin{minipage}[b]{0.49\textwidth}
    \includegraphics[width=3.4in]{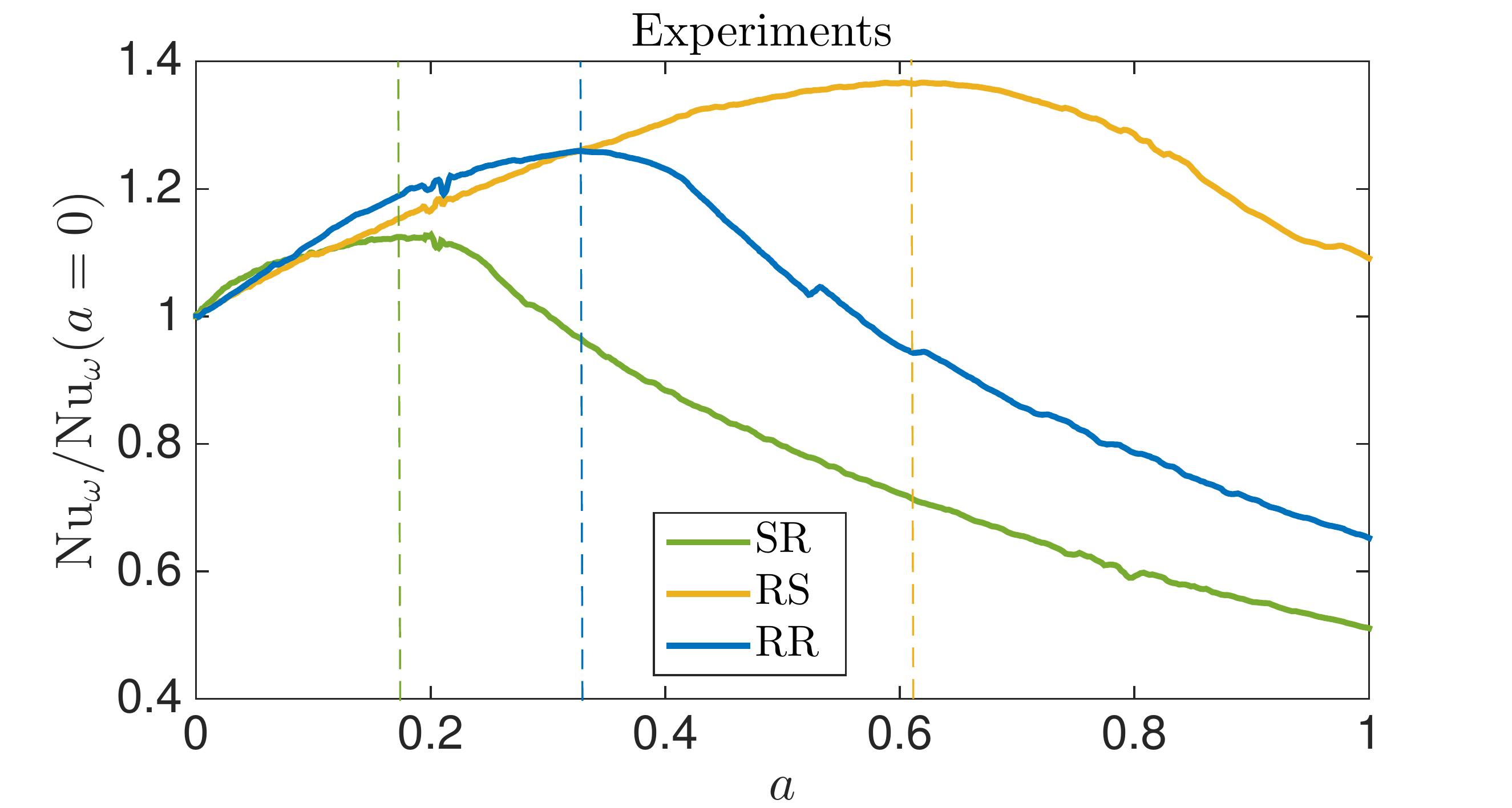}
     \end{minipage}
\caption{{\bf Optimal transport peaks in the case of $h=0.05d$ for 6 ribs.} This figure shows $\Nu$ as function of $a$ for constant driving strength, normalized by its value for $a=0$. Left panel, DNSs with $\Ta=1\times 10^9$. The optimal transport peaks are located at $a_{opt,SR}^{DNS}=0.15\pm0.03$, $a_{opt,RS}^{DNS}=0.55\pm0.04$ and $a_{opt,RR}^{DNS}=0.28\pm0.03$. Right panel,  Experiments with $\Ta=4\times 10^{11} $. The optimal transport peaks are located at $a_{opt,SR}^{EXP}=0.17$, $a_{opt,RS}^{EXP}=0.61$ and $a_{opt,RR}^{EXP}=0.33$. All optimal transport peaks are indicated by the dashed lines, with the respective colors. This figure must be contrasted with Fig.\ \ref{fig:Nu_a}, where the roughness height is higher ($h=0.075d$). Similarly to Fig.\ \ref{fig:Nu_a}, we see the same shift trend of the optimal transport. However, the peak values are different. As expected, when the roughness height is smaller, $a_{opt}$ for SR and RS cases are closer to $a_{opt}$ for the SS case. On the contrary, when the roughness height is larger, $a_{opt}$ for the SR and RS cases deviates more from $a_{opt}$ for the SS case. This can also be clearly seen from Fig.\ \ref{fig:fig10}.}
\label{fig:fig9}

\end{figure}

\begin{figure}[!h]
\begin{center}
\includegraphics[width=3.5in]{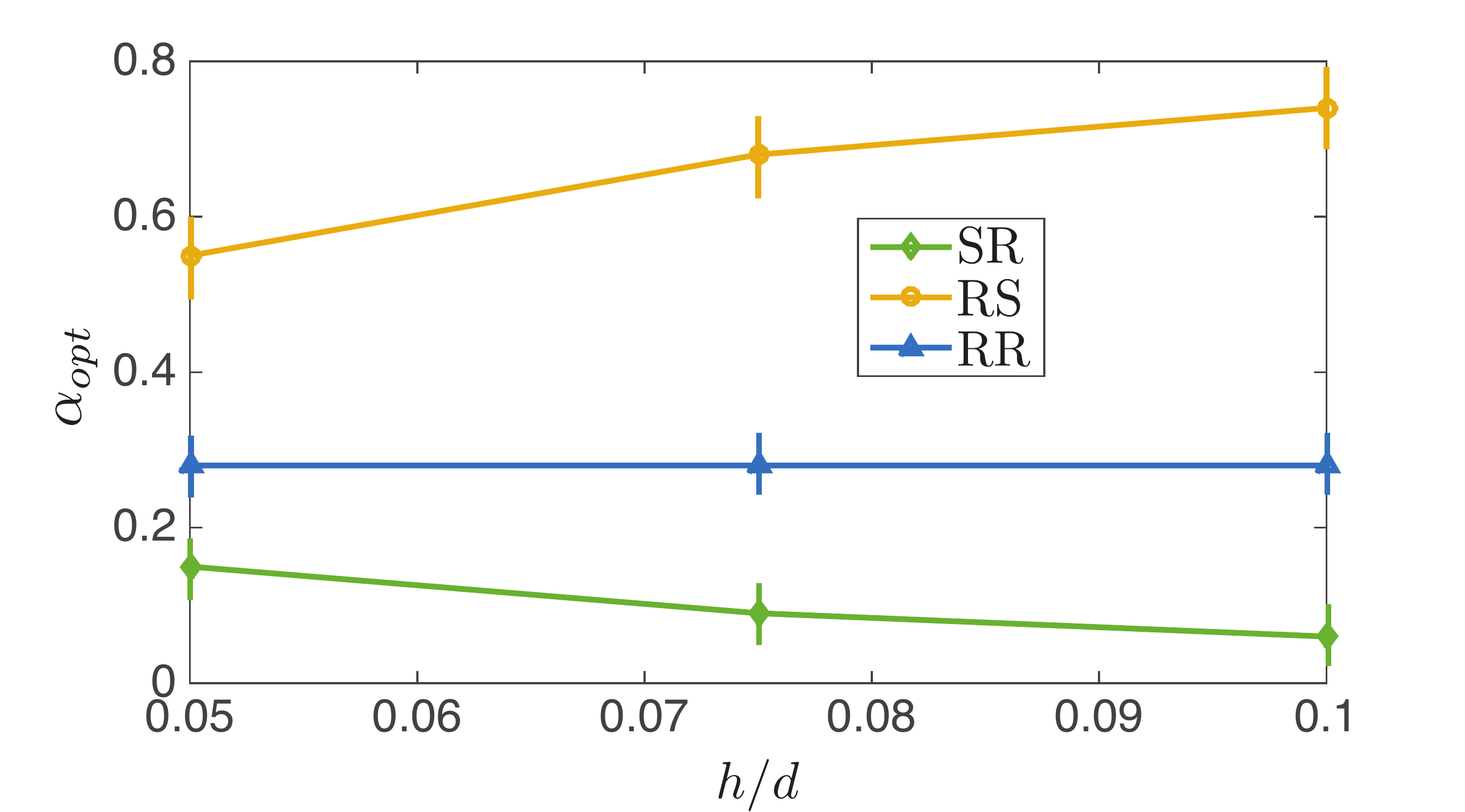}
\caption{ 
{\bf Optimal transport dependence on the roughness height.} Optimal transport rotation ratio $a_{opt}$ as a function of the roughness elements heights $h/d$ for the SR, RS and RR cases, from DNS results. 
}
\label{fig:fig10}
\end{center}
\end{figure}

\begin{figure}
  \includegraphics[width=3in]{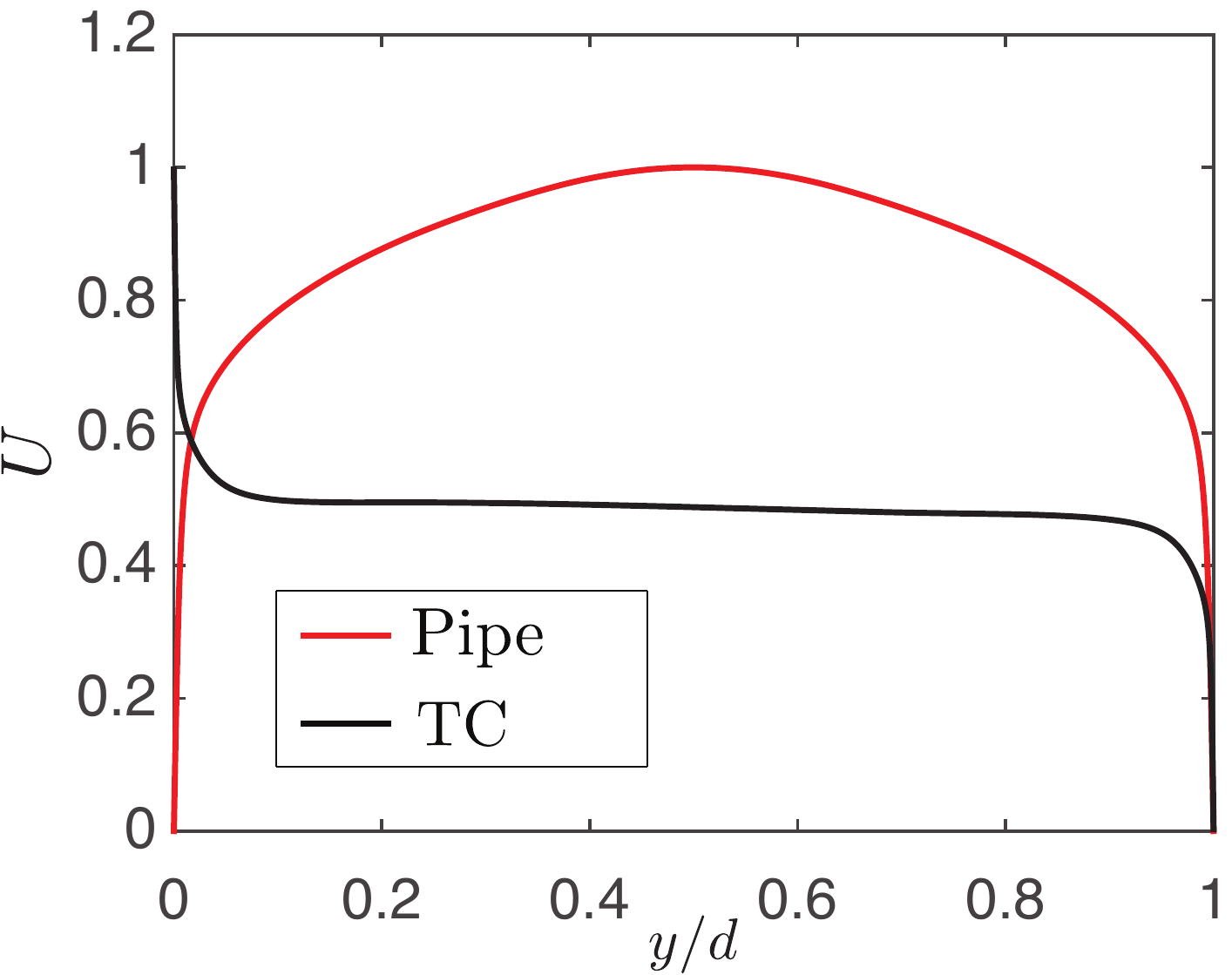}
    \includegraphics[width=3in]{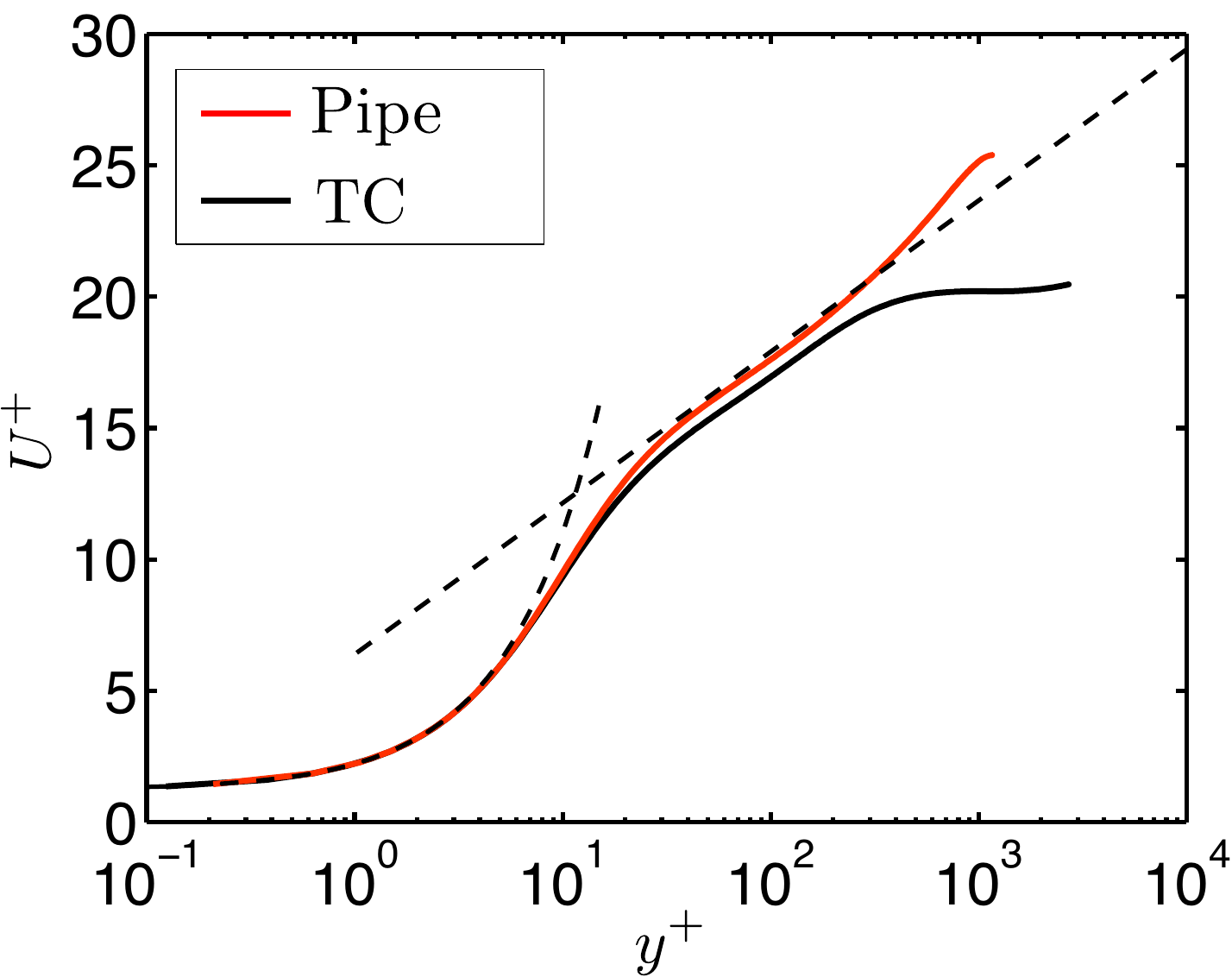}

      \caption{Velocity profiles comparison between pipe and TC flows. Left panel: lengths measured in terms of the gap width of TC resp. the pipe diameter and velocities in terms of the inner cylinder of TC resp. maximal velocity in pipe. Right panel: wall units. Pipe profile is from DNS at $Re_\tau=1000$ \cite{eik13} and TC profile from DNS at $Re_\tau=2000$ at a radius ratio $\eta=0.909$ \cite{ost16jfm}. In the middle of the gap, the velocity profile is much flatter in TC flow than in pipe flow.}
\label{fig:fig12}
\end{figure}

The experiments were performed in the Twente Turbulent Taylor-Couette facility (T$^3$C) \cite{gil11a}, consisting of two independently rotating concentric cylinders. The setup has an inner cylinder with a radius of $r_i=$ 200 mm and an outer cylinder with a radius of $r_o=$ 279.4 mm, resulting in a radius ratio of $\eta = r_i/r_o = 0.716$ and a gap width of $d=r_o-r_i=$ 79 mm. The gap is filled with water with a temperature of T $ \approx 20 ^{\circ}$C. In this work, the inner and outer cylinder rotate up to $\omega_i/2\pi=$ 7.5 Hz and $\omega_o/2\pi=$ 5 Hz, respectively, resulting in Reynolds numbers up to $\text{Re}_i = \omega_i r_i d/\nu = 7.5 \times 10^5$ and $\text{Re}_o= \omega_o r_o d/\nu = 7 \times 10^5$. The cylinders have a height of $L =$ 927 mm, resulting in an aspect ratio of $\Gamma = L/(r_o-r_i) = 11.7$. The end plates rotate with the outer cylinder.
The cylinders were made rough by attaching 2, 3, or 6 vertical strips with a square cross-section (four roughness heights: $2 \times 2$ mm, i.e.\ 2.5\% of the gap width, $4 \times 4$ mm, i.e.\ 5\% of the gap width, $6 \times 6$ mm, i.e.\ 7.5\% of the gap width, and $8 \times 8$ mm, i.e.\ 10\% of the gap width) over the entire height on none, both or either one of the cylinders, similar as in Ref.\ \cite{ber03} (Fig.\ \ref{fig:setup}). The roughness height is larger than the boundary layer thickness \cite{hui13}. 

\subsubsection{Torque measurements}
The torque is measured with a co-axial torque transducer (Honeywell 2404-5K, maximum capacity of 565 Nm), located inside the inner cylinder, to avoid measurement errors due to seals- and bearing friction, as shown in Fig.\ \ref{fig:setup}. In previous studies using this setup, the inner cylinder consisted of 3 different compartments, in which torque was measured in the middle section to exclude end plate effects \cite{gil11,gil12,hui14}. Here, we measure over the entire height of the cylinder, which accounts for the slightly different results for the SS case as compared to these studies.

\subsubsection{Velocity measurements}
Planar particle image velocimetry (PIV) measurements were performed in the $\theta-r$ plane at mid-height ($z=L/2$). We used a high-resolution sCMOS camera (pco.edge camera with 2560 px $\times$ 2160 px resolution), which was operated in double frame mode, as depicted in Fig.\ \ref{fig:setup}. We recorded images through transparent windows in the bottom plate. The flow was illuminated from the side with a pulsed laser (532 nm Quantel Evergreen 145 Nd:YLF). The water was seeded with 20 $\mu$m fluorescent polymer particles (PMMA-RhB-10 by Dantec). The sheet thickness was approximately 1 mm. The PIV measurements were processed using an iterative multi-pass method with final interrogation windows of 32 $\times$ 32 pixel with 50\% overlap and averaged over 500 image pairs per measurement. This results in the averaged azimuthal velocity profile $\langle u_{\theta}(r) \rangle$.

\subsection{Numerical methods}
The motion of the fluid is governed by the incompressible Navier-Stokes equations in the frame co-rotating with the outer cylinder
\begin{eqnarray}
\label{equ1}
\frac{\partial \bf{u}}{\partial t}+{\bf{u}} \cdot \nabla {\bf u} &=&- \nabla p + \frac{f(\eta)}{\Ta^{1/2}}{\nabla}^2 {\bf u}-\Ro^{-1}{\bf e }_z\times {\bf u }, \\
\nabla  \cdot {\bf u} &=& 0, 
\end{eqnarray}
where $\bf u$ and $p$ are the fluid velocity and pressure, respectively. $f(\eta)$ is a geometrical factor which has the form 
\begin{eqnarray}
f(\eta)= \frac{(1+\eta)^3}{8\eta^2}.
\end{eqnarray}
$\Ta$ is the Taylor number and $\Ro$ the Rossby number which characterizes the strength of the driving force. The rotation ratio $a=-\omega_o/\omega_i$ can alternatively be expressed as Rossby number
\begin{eqnarray}
\Ro^{-1} =\frac{2\omega_o d}{|\omega_i-\omega_0|r_i}=-2\frac{1-\eta}{\eta}\frac{a}{|1+a|}.
\end{eqnarray}
The inner cylinder Reynolds number $\Re_i=r_i\omega d/\nu$ and outer cylinder Reynolds number $\Re_o=r_o\omega d/\nu$ are associated with $\Ta$ and $\Ro$ through
\begin{eqnarray}
\Re_i=\frac{\Ta^{1/2}} {f(\eta)} \left (1+\frac{\eta \Ro^{-1}}{2(1-\eta)}\right)
\end{eqnarray}
and
\begin{eqnarray}
\Re_o=\frac{\Ro^{-1}\Ta^{1/2}}{2f(\eta)(1-\eta)}.
\end{eqnarray}

The governing equations are solved using an energy conserving second-order finite-difference code \citep{ver96}, in combination with an immersed-boundary method \citep{fad00,yan06} to deal with the roughness. To achieve high performance computation, a two-dimensional MPI decomposition technique (MPI-pencil) \citep{poe15cf} is adopted. Weak and strong scaling tests show the linear behaviour of the code up to 64K cores. 
The code has been extensively validated and used for TC flow with smooth \cite{ost14pof, ost14pd, ost15pof} and rough \cite{zhu16,zhu17} walls. The axial direction is periodic and thus the end plate effects \cite{avi12} are eliminated. The radius ratio is chosen as $\eta=0.716$. The aspect ratio of the computational domain $\Gamma=L/d$, where $L$ is the axial periodicity length, is taken as $\Gamma=2.09$. The ribs are equi-distributed in the azimuthal
direction, similarly to the experimental implementation (with one more roughness height at 1.5\% of the gap width). The computation box is tested to be large enough to capture the sign changes of the azimuthal velocity autocorrelation at the mid-gap, as suggested as a criterion for the box size \cite{ost15pof}. An appropriate number of grid points
is chosen to make sure that enough resolution has been employed \cite{ost14pof,ost14pd}. E.g.\ at $\Ta=2.15\times10^9$ for the RR case with 6 ribs at roughness height 10\% of the gap width, $3072\times1536\times1536$ grid points are used.

\subsection{Extention of the Grossmann-Lohse theory to the case with wall roughness}

To explain the asymptotic ultimate scaling $1/2$ found in this manuscript, we first recall the origin of the logarithmic correction. We take the only-inner-rotation-case as an example. According to the extension of the Grossmann-Lohse (GL) theory to the ultimate regime\cite{gro11}, the local dissipation rate in the turbulent boundary layer \cite{ll87} can be approximated by 
\begin{eqnarray}\label{EQ10}
\epsilon_u(y)=u_\tau^3/(\kappa y), 
\end{eqnarray}
where $u_\tau=\sqrt  {\tau / (2\rho \pi r^2 L) }$ is the friction velocity, with $\rho$ the fluid density, $\kappa$ the von K\'arm\'an constant. The radius $r$ can be either the inner cylinder radius $r_i$ or the outer one $r_o$, and $y$ the distance from the wall. $u_\tau$ is connected with the inner cylinder velocity $U=r_i\omega_i$ through the law of the wall \cite{sch00}, which is shown for TC turbulence in Refs. \cite{hui13,ost14pof} to obey 
\begin{eqnarray}\label{EQ3}
\frac{u_\tau}{U}=\frac{\kappa}{\mathrm{ln}({B} \Re_i   {u_\tau}/{U} )}.
\end{eqnarray}
$\Re_i$ is the inner cylinder Reynolds number and which for pure inner cylinder rotation can be related to $\Ta$ through the expression $\Ta=\frac{(1+\eta)^6}{64\eta^4}\Re_i^2$, and $B$ is a constant depending on the system geometry. By averaging the local dissipation rate along the radius, we can estimate the mean dissipation rate as
\begin{eqnarray}\label{EQ4}
\epsilon_{u,m} &\sim& \frac{1}{d/2} \int_{0}^{d/2} \epsilon_u(y)dy    \nonumber\\
                       &=& \nu^3d^{-4}\Re_i^3 {\cal{L}}(\Re_i)                                                                             \nonumber\\
                       &=& \nu^3d^{-4}\Re_i^3 \left( \frac{u_\tau}{U} \right)^3 \frac{2}{\kappa}\mathrm{ln}\left(\Re_i\frac{u_\tau}{U}\frac{1}{2}\right).
\end{eqnarray}
Here we assume that logarithmic boundary layer extends from the wall to the mid-gap. Usually how far the log-layer extends depends on $\Re_i$ and can be a small fraction of the gap width, but still for both TC and pipe flows, taking the half gap width or radius  is a reasonable approximation to derive the friction laws \cite{lat92,lew99,sch00,pop00}. The term ${\cal{L}}(\Re_i)= (u_\tau/U)^3 \mathrm{ln}(\Re_iu_\tau/U)$, depending on $\Re_i$, is the logarithmic correction \cite{gro11}. Using the well known exact relation between $\epsilon_{u,m}$ and $\Nu$, namely 
\begin{eqnarray}
\epsilon_{u,m}=\nu^3 d^{-4}\Ta(\Nu-1)\left (\frac{\sqrt \eta}{(1+\eta)/2} \right)^8
\end{eqnarray}
 \cite{eck07b} and with $\Ta\sim  \Re_i^2$, one obtains
 \begin{eqnarray}\label{EQ5}
\frac{\epsilon_{u,m}}{\nu^3d^{-4}} \sim \Re_i^3 {\cal{L}}(\Re_i)  \,\, \mathrm{and} \,\, \Nu \sim \Ta^{1/2}\ {\cal{L}}(\Re_i).                                                                       
\end{eqnarray}
with the logarithmic correction ${\cal{L}}(\Re_i)$ for both dissipation rate and torque scalings. It leads to a less steep increase of $\epsilon_u$ with increasing $\Re_i$ than in the Kolmogorov bulk which scales as $\Re_i^3$, and hence decreases the torque scaling between $\Nu$ and $\Ta$ from the asymptotic ultimate scaling $1/2$ to the effective scaling 0.38 \cite{gro11,gil11,he12,gro16}, as mentioned before.  

With both walls roughened, the log-law in the fully rough regime ($u_\tau h/\nu>70$ \cite{sch00}; all our rough cases are in this regime) becomes
 \begin{eqnarray}\label{EQ6}
\frac{u_\tau}{U}=\frac{\kappa}{\mathrm{ln}(B d/h)},                                                                    
\end{eqnarray}
as shown for turbulent TC flow with one rough boundary layer in Ref. \cite{zhu17}.  The momentum transfer between the wall and the fluid is accomplished by the shear, which in the fully rough regime occurs predominantly by the pressure forces on the side surfaces of the rough elements, rather than by viscous forces \cite{pop00}. That in the ultimate regime the kinematic viscosity $\nu$ is an irrelevant parameter, is reflected in the velocity profile (Eq.~(\ref{EQ6})) being  {\it{independent}} of $\Re_i$. Replacing the velocity profile from the smooth one to the rough one in Eqs.~(\ref{EQ10}, \ref{EQ3}, \ref{EQ4}), remarkably we find that the logarithmic correction term for $\epsilon_{u,m}$ turns into a \textit{constant} and thus its effect on the scaling exponent vanishes. The mean dissipation rate and torque thus now scale as
 \begin{eqnarray}\label{EQ7}
\frac{\epsilon_{u,m}}{\nu^3d^{-4}} \sim \Re_i^3  \,\, \mathrm{and} \,\, \Nu \sim \Ta^{1/2},                                                                       
\end{eqnarray}     
which explains the asymptotic ultimate regime scaling seen in Fig.~\ref{fig:fig2} for the RR case. In the RS or SR case, the boundary layer at the smooth wall depends on $\Re_i$ while the boundary layer at the rough wall is independent of it. Therefore, in these cases the logarithmic correction is reduced but not totally canceled.



\bibliographystyle{prsty_withtitle} 

\bibliography{literatur}



\newpage

\noindent {\bf
Acknowledgements} 

We gratefully acknowledge V. Mathai for insightful discussions. We would like to thank G. W. Bruggert and M. Bos, as well as G. Mentink and R. Nauta for their technical support and D.P.M. van Gils and R. Ezeta for various discussions and help with the experiments. The work is financially supported by the Dutch Foundation for
Fundamental Research on Matter (FOM), the Netherlands Center for Multiscale Catalytic
Energy Conversion (MCEC), the Dutch Technology Foundation (STW) and a VIDI grant (No.\ 13477), all sponsored by the Netherlands
Organisation for Scientific Research (NWO).  We thank the Dutch Supercomputing Consortium
SURFsara, the Italian supercomputer Marconi-CINECA through the PRACE Project No. 2016143351 and the ARCHER UK National Supercomputing Service through the DECI Project 13DECI0246 for the allocation of computing time.
\\

\noindent{\bf
Author Contributions}
X.Z., S.G.H., R.A.V., R.V., C.S. and D.L. conceived the ideas.
R.A.V. and D.B. performed the measurements. X.Z. performed the numerical simulations. X.Z. and R.A.V. analyzed the data. 
X.Z., R.A.V. and D.L. wrote the paper. 
R.V., C.S. and D.L. supervised the project. 
 All authors discussed the physics and proofread the paper. \\

\noindent{\bf
Author Information}
Reprints and permissions information is available at
www.nature.com/reprints. The authors declare no competing financial
interests. Readers are welcome to comment on the online version of the paper. 
Correspondence and requests for materials should be addressed to
D.L. (d.lohse@utwente.nl) or C.S. (chaosun@tsinghua.edu.cn).


\newpage

\section{Extended Data}


\end{document}